\documentclass[lettersize,journal]{IEEEtran}

\IEEEoverridecommandlockouts
\usepackage{amsmath,amsfonts}
\usepackage{algorithmic}
\usepackage{algorithm}
\usepackage{array}
\usepackage[caption=false,font=footnotesize]{subfig}
\usepackage{textcomp}
\usepackage{stfloats}
\usepackage{url}
\usepackage{verbatim}
\usepackage{graphicx}
\usepackage{cite}
\usepackage{multirow}
\usepackage{epstopdf}
\usepackage{longtable}
\usepackage{setspace}
\usepackage{bm}
\usepackage{tabularx}
\usepackage{makecell}
\usepackage{tikz}

\usetikzlibrary{arrows.meta,positioning,shapes.geometric,calc}

\def\BibTeX{{\rm B\kern-.05em{\sc i\kern-.025em b}\kern-.08em
    T\kern-.1667em\lower.7ex\hbox{E}\kern-.125emX}}

\ifCLASSINFOpdf

\else

\fi

\begin{document}

\renewcommand{\baselinestretch}{1.00}

\title{Mobility-Aware Task Offloading in Mobile Edge Computing-Empowered Space-Air-Ground Integrated Internet-of-Things Networks}

\vspace{-2.8cm}

\author{Minghao~Sun,
        Jinghuai~Xiong,
        Xiaoli~Chu,~\IEEEmembership{Senior Member, IEEE,}
        Di~Yuan,~\IEEEmembership{Senior Member, IEEE,}
        Zitian~Zhang
\thanks{Minghao Sun and Xiaoli Chu are with the School of Electrical and Electronic Engineering, the University of Sheffield, UK (e-mail:\{msun39, x.chu\}@sheffield.ac.uk).
Jinghuai Xiong is with the College of Computing and Data Science, the Nanyang Technological University, Singapore (e-mail: JINGHUAI001@e.ntu.edu.sg).
Di Yuan is with the Department of Information Technology, Uppsala University, Sweden (e-mail: di.yuan@it.uu.se).
Zitian~Zhang is with the School of Information and Electronic Engineering, Zhejiang Gongshang University, China (e-mail: zitian.zhang@mail.zjgsu.edu.cn).
(Corresponding author: Zitian Zhang)}
\thanks{This work was supported in part by the Horizon Europe Research and Innovation Program under grant 101086219, the UK EPSRC grant EP/X038971/1, Royal Society International Exchanges 2024 Cost Share (IEC$\setminus$NSFC$\setminus$242607), National Natural Science Foundation of China (62301488, W2421086, 62401506), and Zhejiang Provincial Science and Technology Innovation Key Project (2023R5211).
}

}

\markboth{IEEE submitted manuscript}%
{Submitted paper}

\maketitle

\renewcommand{\baselinestretch}{1.00}
\vspace{-0.8cm}
\begin{abstract}
Mobile edge computing (MEC)-empowered space-air-ground integrated Internet-of-Things (IoT) networks (SAGIINs) provide a promising solution for computation-intensive tasks generated by mobile IoT devices (IDs). 
However, task offloading in SAGIINs is challenging due to time-varying ID-to-unmanned aerial vehicle (UAV) channels, intermittent LEO satellite service availability, and the coupling among UAV trajectory, communication, computation, and energy decisions.
This paper investigates an energy-efficient task offloading and UAV trajectory planning problem in an MEC-empowered SAGIIN, where one UAV acts as both an aerial relay and an MEC server, and multiple LEO satellites provide supplementary computing resources. The UAV trajectory planning, task splitting, and computing resource allocation are jointly optimized to minimize the total system energy consumption under task completion, communication, computation, flight region, and satellite service window constraints. 
The resulting non-convex problem is decomposed into a UAV trajectory planning subproblem and a task splitting and computing resource allocation subproblem.
To solve the former, a retrieval-augmented generation (RAG)-assisted large language model (LLM)-enhanced deep deterministic policy gradient (DDPG) algorithm, termed RLD, is proposed. 
The latter is shown to be convex and can be solved by using CVXPY for a given UAV trajectory. A weighted retrieval score is further designed to balance ID mobility-pattern similarity and historical UAV trajectory energy performance. 
Simulation results show that RLD completes all computational tasks in previously unseen test scenarios while consuming less total energy than the benchmark algorithms.





\end{abstract} 

\begin{IEEEkeywords}Internet of Things, satellite mobile edge computing, space-air-ground integrated networks, unmanned aerial vehicle, large language model, reinforcement learning.
\end{IEEEkeywords}

\section{Introduction}
\IEEEPARstart{T}{he} rapid development of the Internet of Things (IoT) is enabling a wide range of intelligent and mobile applications, such as intelligent transportation, public safety monitoring, emergency response, low-altitude logistics, and mobile robotic services \cite{SIoT_survey1}. 
These applications usually require massive data collection, timely task processing, and reliable connectivity \cite{IoT_survey}.
However, the limited computational capability of IoT devices (IDs) prevents them from executing computation-intensive tasks locally.
Low-Earth-orbit (LEO) satellite-assisted mobile edge computing (MEC) systems have been widely regarded as a promising solution for providing high-performance computing services and complementing the limited computational resources of terrestrial IoT networks \cite{ SIoT_survey2}. 
In such systems, the data collected by IDs is typically transmitted to cloud, edge, or satellite computing nodes for further processing. 

However, direct task offloading from IDs to LEO satellites may suffer from long propagation distances, limited link capacity, and intermittent service availability caused by satellite mobility. 
To overcome these limitations, space-air-ground integrated IoT networks (SAGIINs) provide a promising architecture for such scenarios \cite{SEC_survey}.
Different from traditional networks that mainly rely on fixed terrestrial infrastructure, SAGIINs can integrate terrestrial IDs, unmanned aerial vehicles (UAVs), and LEO satellites to provide flexible communication, computing, and relaying services. 
In this architecture, UAVs can be employed as aerial access points, lightweight MEC servers, or aerial relays \cite{UAV_survey}. 
Leveraging their controllable three-dimensional mobility, UAVs can be rapidly deployed and dynamically repositioned to accommodate the spatial distribution and service demands of IDs by creating line-of-sight air-ground links. 
Meanwhile, LEO satellites can provide wide-area coverage, supplementary computing resources, and resilient backhaul support when terrestrial or aerial resources are overloaded, unavailable, or insufficient.

Deep reinforcement learning (DRL) has been widely used for dynamic resource allocation and sequential decision-making in satellite-assisted MEC systems \cite{SEC_RL_1, SEC_RL_2, SEC_RL_3, SEC_RL_4}. 
Among various DRL algorithms, deep deterministic policy gradient (DDPG) is suitable for continuous control problems and can be applied to UAV trajectory planning and continuous resource allocation in MEC-empowered SAGIINs \cite{DDPG_MEC}.
Nevertheless, conventional DRL methods often suffer from poor initial performance, low exploration efficiency, slow convergence, and unstable training in large and constrained action spaces. 
Moreover, existing studies above focus on relatively simple or limited mobility scenarios, which may restrict the adaptability of the learned policies to dynamic environments. 
In contrast, SAGIINs involve highly time-varying network conditions due to the mobility of IDs, UAVs, and LEO satellites, requiring fast and adaptive decision-making. Therefore, an efficient offline training and online decision-making framework is needed to generate feasible solutions with low online decision latency.

To overcome these limitations, large language models (LLMs) can be used as an auxiliary source of prior knowledge for DRL-based optimization. 
LLMs have demonstrated strong capabilities in semantic understanding, reasoning, and human-like language interaction \cite{LLM}.  
This capability is particularly useful for DRL-based network optimization, where a well-designed initial policy can reduce inefficient exploration and improve convergence performance \cite{LLM_RL_CS_1}.
However, the mobility of IDs, UAVs, and LEO satellites and their interplay have not been sufficiently studied in existing works on LLM and DRL-enhanced SAGIINs.

To address the above challenges, this paper proposes a retrieval-augmented generation (RAG)-assisted LLM-enhanced DDPG (RLD) framework for mobility-aware task offloading and resource allocation in MEC-empowered SAGIINs.

The main contributions of this paper are summarized as follows:
\begin{itemize}
\item 
Unlike conventional SAGIIN models that assume static ground users and continuously available satellite services, we develop a mobility-aware, highly time-varying MEC-empowered SAGIIN model consisting of multiple moving IDs, one UAV, and multiple LEO satellites.
The proposed model explicitly captures the continuous movement of ground IDs and the time-varying service windows of LEO satellites. 
Based on this model, we formulate an energy-efficient task offloading and UAV trajectory planning problem that minimizes the total communication, computing, and UAV propulsion energy consumption subject to task-completion constraints, UAV and satellite computing-capacity constraints, ID-to-UAV and UAV-to-satellite transmission-capacity constraints, UAV flight-region constraints, and time-varying satellite service-window constraints.
Due to the coupling among these variables, the problem is decomposed into trajectory planning and resource allocation subproblems. 
The trajectory subproblem is non-convex because of the rotary-wing propulsion model and trajectory-dependent links, whereas the task-splitting and resource-allocation subproblem becomes convex for a given UAV trajectory and is solved using CVXPY.

\item 
To solve the UAV trajectory planning subproblem, we propose a novel RLD algorithm, wherein a retrieval score is introduced to enable the RAG module to improve the exploration of DDPG for UAV trajectory planning.
The RLD model is trained offline and can generate UAV trajectories during online decision-making across different scenarios. 
To improve exploration efficiency and generalization, an energy-aware retrieval score is designed for the RAG-LLM module, which jointly considers the similarity between the current and historical ID mobility patterns and the energy consumption of the corresponding historical UAV trajectories, with a weight factor used to balance their contributions.

\item Simulation results show that the RLD algorithm completes all computational tasks and reduces the total system energy consumption compared with benchmark algorithms across all considered testing scenarios, demonstrating its effectiveness and adaptability in dynamic SAGIIN environments.
Moreover, the weight factor analysis confirms the effectiveness of jointly considering mobility-pattern similarity and historical energy performance during case retrieval in the RAG-LLM module.
\end{itemize}

The remainder of this paper is organized as follows. Section II reviews related works on satellite-assisted MEC, DRL-based resource allocation, and LLM-assisted learning methods. Section III presents the system model, including communication, computation, mobility, and energy consumption models. Section IV formulates the energy-efficient task offloading and resource allocation problem. Section V introduces the proposed RLD framework and its learning procedure. Section VI presents the simulation results and performance analysis. Finally, Section VII concludes this paper.


\section{Related Work}

Existing studies have investigated task offloading and resource allocation in satellite edge computing from different perspectives, including model-driven optimization methods \cite{SAG_close, SEC_alg_1, SEC_alg_2, SEC_alg_3, SEC_alg_4}, game-theoretic approaches \cite{SEC_game_1, SEC_game_2}, and heuristic algorithms \cite{SEC_GA_1, SAG_heuristic}. These conventional methods usually rely on accurate mathematical structures, fixed network assumptions, or repeated problem decomposition. 
For example, \cite{SAG_close} provided an effective model-driven solution for multi-UAV-enabled SAGIIN by jointly optimizing mobile user-UAV association, UAV trajectory, task offloading, transmit power, and computing frequency. 
However, this method may suffer from high computational complexity. 
In addition, \cite{SEC_alg_4} analyzed an IoT-supportable satellite edge computing architecture and investigated network slice scheduling, where latency, computational power, and transmission power were jointly optimized to meet different IoT service requirements. However, only a limited set of decision variables was considered, which restricted its ability to model the complex coupling between communication and computation in practical SAGIINs. 

Heuristic algorithms and decomposition-based approaches have also been investigated to reduce computational complexity. In \cite{SEC_GA_1}, QoS-aware multi-hop task offloading in satellite-terrestrial edge networks was studied, and a hybrid algorithm combining genetic search and the Lagrangian multiplier method was proposed to minimize the total task latency while satisfying QoS constraints. 
The authors of \cite{SAG_heuristic} studied LEO-satellite-assisted UAV data collection, where UAVs collected data from remote IDs and transmitted it through LEO satellite relays. They jointly optimized UAV trajectory and data transmission to reduce UAV energy consumption using Dantzig-Wolfe decomposition, column generation, and heuristic algorithms. 
Although these studies provide useful model-driven and heuristic solutions, they are mainly based on static system models, discretized trajectory formulations, or limited decision variables. 
Therefore, they cannot fully capture the multi-slot dynamic decision-making process, time-varying satellite service windows, UAV trajectory-dependent channels, or coupled communication-computation-energy constraints in practical SAGIIN scenarios \cite{limit}.

Learning-based methods have also been introduced to improve the adaptability of task offloading and resource allocation in dynamic MEC systems.
Authors in \cite{SAG_drl_1} proposed a DRL-based method for LEO-MEC-assisted energy-harvesting UAV systems to jointly optimize satellite selection, transmission power, and partial task offloading, where an action shaping mechanism was used to handle the mixed discrete-continuous decision space.
In \cite{SAG_drl_2}, the authors developed a multi-heterogeneous-agent DRL framework with cross-layer state prediction and reward redistribution to improve congestion control under dynamic SAGIIN link conditions and heterogeneous QoS requirements.
In \cite{SEC_RL_1}, the authors investigated task partitioning and resource allocation, where a deep recurrent Q-learning network was adopted to capture temporal features and combined with federated learning to improve decision-making efficiency.
In \cite{SEC_RL_2}, the authors proposed a model-based two-tier RL algorithm to jointly optimize task offloading and resource allocation in satellite-assisted IoT networks.
In \cite{SEC_RL_3}, the authors decomposed the long-term stochastic computation offloading problem into deterministic per-slot subproblems and combined DRL with traditional optimization algorithms.
In \cite {SEC_RL_4}, the authors studied task offloading service migration in satellite edge computing and developed a DRL-based method to optimize offloading path selection under dynamic service demands and constrained bandwidth, energy, and storage resources.
However, these DRL methods may suffer from poor initial performance, inefficient exploration, slow convergence, and unstable training when facing large, constrained, and highly coupled action spaces.

Recently, LLMs have shown strong potential in semantic understanding, reasoning, and decision assistance, which provides a new opportunity for enhancing DRL-based network optimization \cite{LLM_RL_CS_9}.
Existing works have shown that LLMs can reduce the sample complexity of RL in robotic manipulation \cite{LLM_RL_CS_6}, guide exploration through linguistic priors and inductive biases \cite{LLM_RL_CS_7}, generate meaningful goals for exploration \cite{LLM_RL_CS_13}, and enhance decision-making in recommender systems, code optimization, and information extraction tasks \cite{LLM_RL_CS_8,LLM_RL_CS_12,LLM_RL_CS_1}.
In wireless communication networks, LLMs have also been used to analyze user objectives and constraints, select suitable DRL experts, assist RL-based network deployment, and improve RL training through curriculum generation and fine-tuning strategies \cite{LLM_RL_WL_1,LLM_RL_WL_2,LLM_RL_WL_4}.

Overall, most existing studies focus on either satellite-assisted or UAV-assisted MEC systems. 
Although SAGIINs have been investigated in recent works, the joint modeling of UAV trajectory control, satellite service windows, parallel task splitting, and communication-computation-energy coupling is still limited. 
More importantly, the integrated mobility of IDs, UAVs, and LEO satellites has not been sufficiently addressed. 
This limitation becomes more critical in emerging scenarios such as the low-altitude economy, where the movement of ground devices, aerial platforms, and satellites can jointly reshape the communication links, computing opportunities, and service availability.

\section{System Model}
\subsection{System Overview}

\begin{figure}
    \centering
    \includegraphics[width=1\linewidth]{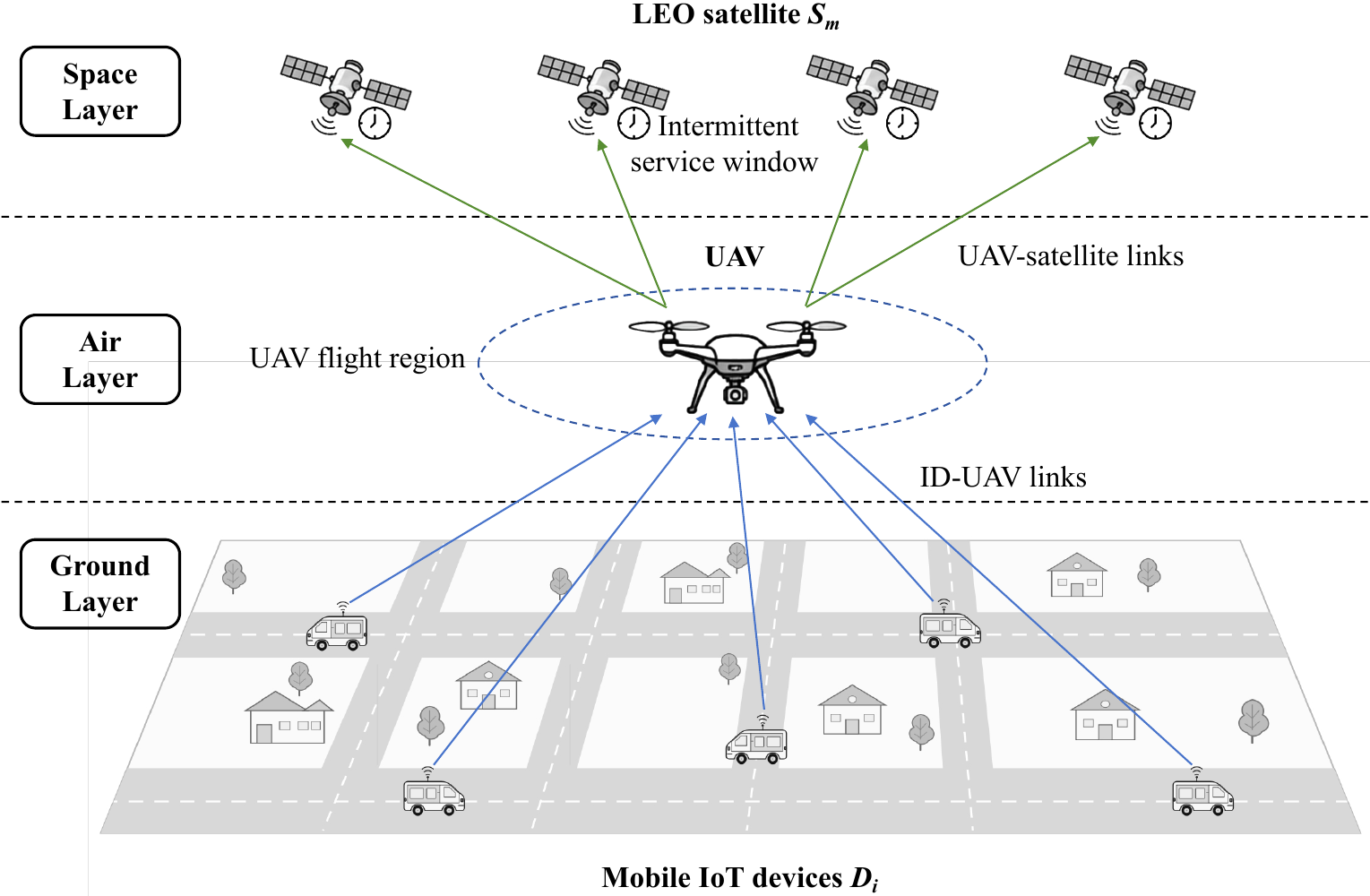}
    \caption{The system model of the mobility-aware MEC-empowered SAGIIN.}
    \label{system}
\end{figure}


The MEC-empowered SAGIIN considered in this work is shown in Fig. 1. 
The system operates in scheduling periods, each divided into $T$ time slots, denoted by $\mathcal{T}=\{1, 2,\ldots, T\}.$
Each time slot $t \in \mathcal{T}$ has a fixed duration of $\Delta$ (in seconds).
There are $I$ mobile IDs distributed across a suburban area, such as unmanned delivery vehicles used for delivery or sensing services.
We label the IDs as ${\cal {D}}=\{D_1,D_2,\ldots,D_I\}$ and assume that each $D_i \in {\cal {D}}$ moves along a predetermined trajectory on the horizontal $xy$-plane. 
Specifically, the location of $D_i$ in time slot $t$ is denoted by $(X^{(D)}_i[t],Y^{(D)}_i[t],0)$.  
At the beginning of a scheduling period, each ID will generate one computational task. The computational task generated by ID $D_i$ is denoted by $\{V^{(D)}_i, \eta^{(D)}\}$, where $V_i^{(D)}$ denotes the input data size in bits and $\eta^{(D)}$ denotes the required number of CPU cycles per bit.
To reduce weight and energy consumption, the IDs are not equipped with onboard computing resources. Instead, their computational tasks can be offloaded to the UAV for processing or further relayed to LEO satellites for processing. 
Without loss of generality, we assume that during each scheduling period, there are $M$ LEO satellites, denoted by ${\cal {S}}=\{S_1,\;S_2,\;...\;S_M\}$, providing MEC services to the IDs.
Due to high-speed mobility of LEO satellites, each $S_m \in {\cal {S}}$ has a limited temporal window of $[T^{(S)}_{m,st},T^{(S)}_{m,ed}]$ ($0 \le T^{(S)}_{m,st} < T^{(S)}_{m,ed}$) to serve the IDs when they are in $S_m$'s coverage. 
The service availability of $S_m$ in time slot $t$ is denoted by
\begin{equation}
    \chi_m[t]
    =
    \begin{cases}
    0,
    & t \notin \left[ {T_{m,st}^{(S)},T_{m,ed}^{(S)}} \right], \\[2mm]
1,
    & t \in \left[ {T_{m,st}^{(S)},T_{m,ed}^{(S)}} \right].
    \end{cases}
\end{equation}

If $S_m$ continues to cover the IDs after a scheduling period ends, $T^{(S)}_{m,ed}$ is set to $T$. 

\subsection{Communication and Computation Models}
In this system, the downlink transmission latency is neglected since the computation result size is typically much smaller than the uplink task data size.
For ID-to-UAV communication, we assume that each ID is equipped with a single transmitting antenna and the UAV is equipped with a receiving antenna. 
Each ID is allocated an orthogonal sub-channel with a bandwidth of $B$ and a fixed transmission power $p^{(D)}$.
The UAV can receive data from multiple IDs concurrently. 
For UAV-to-satellite communication, we assume that the UAV sends the task data to each satellite via an orthogonal channel with a bandwidth of $W$.
In the 3-D Cartesian coordinate system, the trajectory of UAV $U$ in a scheduling period is denoted by $tr^{(U)}[t] = [x^{(U)}[t],y^{(U)}[t],Z^{(U)}],-R \leq x^{(U)}[t],y^{(U)}[t] \leq R, t \in \mathcal{T}$, where the UAV has a fixed altitude of $Z^{(U)}$ to ensure safe operation in suburban environments and stable communication performance, while $x^{(U)}[t]$ and $y^{(U)}[t]$ denote the UAV's horizontal position in time slot $t$. 
Here, $R$ denotes the half-length of the square horizontal flight region centered at the origin, with $[0,0, Z^{(U)}]$ representing the UAV's initial horizontal position.
In time slot $t$, the channel power gain between ID $D_i$ and UAV $U$ is given by \cite{UAV_ID_link}

\begin{equation}
		\begin{array}{l}
			G _{i}^{D - U}[t] = 
			\frac{{{\ell _{ref,UAV} \cdot (|h_i^{I-U}[t]|)^2}}}{{{{\left( {X_i^{(D)}[t] - x^{(U)}[t]} \right)}^2} + {{\left( {Y_i^{(D)}[t] - y^{(U)}[t]} \right)}^2} + {{\left( {{Z^{(U)}}} \right)}^2}}},
		\end{array}
		\label{eq:channel power gain}
\end{equation}

\noindent
where ${{\ell _{ref,UAV}}}$ represents the channel power gain at the reference distance of $\Omega_{ref} = 1$m, $h_i^{I-U}[t]$ represents the Rician fading with factor $K_1$.
According to Shannon's theory, the maximum achievable uplink data rate from $D_i$ to $U$ is given by:

\begin{equation}
		\gamma _{i}^{D - U}[t] = B\log \left( {1 + \frac{{ p^{(D)} \cdot G _{i}^{D - U}[t]}}{{\sigma _{SC}^2}}} \right),
		\label{D-U capcacity}
\end{equation}

\noindent
where $\sigma _{SC}^2$ is the additive white Gaussian noise (AWGN) power in a sub-channel.

The UAV-to-satellite channel power gain is modeled as
\begin{equation}
\ell_m^{U-S}[t]
=
G_U G_S
\left(
\frac{\lambda_c}{4\pi d^{U-S}}
\right)^2 (|h_m^{U-S}[t]|)^2,
\label{eq:large_scale_gain_us}
\end{equation}
where $G_U$, $G_S$, $\lambda_c$, $d^{U-S}$, and $h_m^{U-S}[t]$ denote the UAV antenna gain, satellite antenna gain, the carrier wavelength, the distance between the UAV and satellite $S_m$, and the Rician fading with factor $K_2$, respectively. Notably, since the UAV displacement during one scheduling period is much smaller than the UAV-to-satellite link distance, the UAV-to-satellite distance is approximated as constant for tractability.

Therefore, the achievable data rate from
the UAV to satellite $S_m$ in time slot $t$ is given by
\begin{equation}
\gamma_m^{U-S}[t]
=
W\log_2
\left(
1+
\frac{
p^{U-S} \ell_m^{U-S}[t]
}
{\sigma_{\mathrm{US}}^2}
\right),
\label{eq:us_rate_rician}
\end{equation}
where $W$ is the UAV-to-satellite link bandwidth, $p^{U-S}$ is the fixed
UAV transmit power for satellite $S_m$, and $\sigma_{{US}}^2$
is the AWGN power at the satellite receiver.
Since the UAV-to-satellite transmit power has a relatively small impact on the total energy consumption compared with the UAV propulsion energy, $p^{U-S}$ is assumed to be constant for model tractability.

For simplicity, we ignore the propagation delays between IDs and the UAV and assume that a computational task can be processed at an MEC node, i.e., a UAV or satellite, immediately after this task’s first data bit has been offloaded to the MEC node.

Assuming that the computational tasks are splittable, we consider a distributed parallel computing model. 
In this model, task $D_i$ can be split into up to $M+1$ portions and processed at $M+1$ processing nodes concurrently. 
A task is considered complete only when all of its portions have been processed successfully.
To ensure reliable task processing, the computation of each task portion offloaded to an LEO satellite must be completed before this satellite's serving window closes. 
The UAV and each satellite's computational capabilities are denoted by $F^{(U)}$ and $F^{(S)}$ (in CPU cycles/s), respectively. 
Let $\alpha _{i}^{D - U}[t]$ and $\alpha _{i,m}^{D - S}[t]$ denote the portions of task $D_i$ allocated to UAV $U$ and satellite $S_m$ in time slot $t$, respectively, where
\begin{equation}
\sum_{t=1}^{T}
\left(
\alpha_i^{D-U}[t]
+
\sum_{S_m \in \mathcal{S}}
\alpha_{i,m}^{D-S}[t]
\right)
=1,\quad \forall D_i \in \mathcal{D},
\end{equation}
and
\begin{equation}
0 \leq \alpha_i^{D-U}[t],\alpha_{i,m}^{D-S}[t] \leq 1,
\quad \forall D_i \in \mathcal{D}, S_m \in \mathcal{S}, t \in \mathcal{T}.
\end{equation}
Let $f_{i}^{D - U}[t]$ and $f_{i,m}^{D - S}[t]$ (in CPU cycles/s) denote the allocated computational resources of  $U$, and $S_m$ for computational task $D_i$ in time slot $t$, respectively.


\subsection{Energy Consumption Model}
 
In time slot $t$, the transmission energy consumption of $D_i$ over the ID-UAV channel can be expressed as

\begin{small}
	\begin{equation}
		E _{trs,i}^{(D)}[t] = {p^{(D)}} \cdot \Delta.
		\label{eq:eq107}
	\end{equation}
\end{small}

For UAV $U$, its energy consumption contains three parts, i.e., computing energy consumption, transmission energy consumption, and propulsion energy consumption. 
In time slot $t$, the computing energy consumption $E _{cpt,n}^{(U)}[t]$ of UAV $U$ is determined by its CPU rotational speed \cite{energy_equation} as

\begin{small}
	\begin{equation}
		E _{cpt}^{(U)}[t] = \sum\limits_{{D_i} \in {\cal {D}}}{\rho}{\left( { {f_{i}^{D - U}[t]} } \right)^2}\alpha _{i}^{D - U}[t]V_i^{(D)}\eta^{(D)},
		\label{eq:eq108}
	\end{equation}
\end{small}
\vspace{-0.4cm}

\noindent
where $\rho$ is the chip architecture related computing energy factor. 
The transmission energy consumption, $E _{trs}^{(U)}[t]$, can be calculated with its beamforming vectors related to the $M$ satellites as

\begin{small}
	\begin{equation}
		E _{trs}^{(U)}[t] = \sum\limits_{{S_m} \in {\cal {S}}} p^{U-S}\chi_m[t]  \cdot \Delta.
		\label{eq:eq109}
	\end{equation}
\end{small}
\vspace{-0.4cm}






The initial and final positions of the UAV during a scheduling period are denoted by 
${tr}^{(U)}[0]$ and ${tr}^{(U)}[T]$, respectively. 
The velocity of the UAV in time slot $t$ is given by
\begin{equation}
{\nu}^{(U)}[t]
=
\frac{{tr}^{(U)}[t]-{tr}^{(U)}[t-1]}{\Delta},
\quad t\in\mathcal{T}.
\label{eq:uav_velocity}
\end{equation}

Accordingly, the UAV speed is given by
\begin{equation}
V^{(U)}[t]
=
\left\|{\nu}^{(U)}[t]\right\|.
\label{eq:uav_speed}
\end{equation}

For a rotary-wing UAV, the propulsion power consumption in time slot $t$ 
can be expressed as
\begin{equation}
\begin{aligned}
P_{\mathrm{ppl}}^{(U)}[t]
=&\;
P_0
\left(
1+
\frac{3\left(V^{(U)}[t]\right)^2}{U_{\mathrm{tip}}^2}
\right)
\\
&+
P_i
\left(
\sqrt{
1+
\frac{\left(V^{(U)}[t]\right)^4}{4v_0^4}
}
-
\frac{\left(V^{(U)}[t]\right)^2}{2v_0^2}
\right)^{\frac{1}{2}}
\\
&+
\frac{1}{2}d_0\rho_{air} s A
\left(V^{(U)}[t]\right)^3 .
\end{aligned}
\label{eq:uav_propulsion_power}
\end{equation}
where $P_0$ and $P_i$ denote the blade profile power and induced power 
in hovering status, respectively; $U_{\mathrm{tip}}$ is the tip speed of 
the rotor blade; $v_0$ is the mean rotor induced velocity in hovering; 
$d_0$ is the fuselage drag ratio; $\rho_{air}$ denotes the air density; 
$s$ is the rotor solidity; and $A$ is the rotor disc area \cite{UAV_energy}.
Thus, the propulsion energy consumption of the UAV in time slot $t$ is given by
\begin{equation}
E_{\mathrm{ppl}}^{(U)}[t]
=
\Delta P_{\mathrm{ppl}}^{(U)}[t].
\label{eq:uav_propulsion_energy_slot}
\end{equation}

For any satellite $S_m \in {\cal {S}}$, we only consider its computing energy consumption, $E _{cpt,m}^{(S)}[t]$, which is determined by its CPU rotational speed \cite{energy_equation} as

\begin{small}
	\begin{equation}
		E _{cpt,m}^{(S)}[t] = \sum\limits_{{D_i} \in {\cal {D}}}{\rho}{\left( { {f_{i,m}^{D - S}[t]} } \right)^2}\alpha _{i,m}^{D - S}[t]V_i^{(D)}\eta^{(D)},
		\label{eq:eq113}
	\end{equation}
\end{small}

Accordingly, the total system energy consumption over a scheduling period is given by

\begin{small}
	\begin{equation}
		\begin{array}{r}
			E  =\sum\limits_{t = 1}^T { {\left( {E _{cpt}^{(U)}[t] + E _{trs}^{(U)}[t] + E _{ppl}^{(U)}[t]} \right)}}\\
            +\sum\limits_{t = 1}^T {\sum\limits_{{D_i} \in D} {\left( {E _{trs,i}^{(D)}[t]} \right)} }+ \sum\limits_{t = 1}^T {\sum\limits_{{S_m} \in S} {E _{cpt,m}^{(S)}[t]} } .
		\end{array}
		\label{eq:total energy}
	\end{equation}
\end{small}


\section{Problem Formulation and Decomposition}

\subsection{Energy-Efficient Task Offloading and Trajectory Design (EE-TOTD) Problem}
Considering that the IDs, UAV, and satellites have very limited on-board energy reserves, this work aims to minimize the total energy consumption of the system in every scheduling period.

For notational convenience, we collect the time-varying optimization
variables over the entire scheduling period into
\begin{align}
\boldsymbol{\alpha}_{i}^{D-U}
&=
\left[
\alpha_{i}^{D-U}[1],\ldots,\alpha_{i}^{D-U}[T]
\right]^{T},\\
\boldsymbol{\alpha}_{i,m}^{D-S}
&=
\left[
\alpha_{i,m}^{D-S}[1],\ldots,\alpha_{i,m}^{D-S}[T]
\right]^{T},\\
\boldsymbol{f}_{i}^{D-U}
&=
\left[
f_{i}^{D-U}[1],\ldots,f_{i}^{D-U}[T]
\right]^{T},\\
\boldsymbol{f}_{i,m}^{D-S}
&=
\left[
f_{i,m}^{D-S}[1],\ldots,f_{i,m}^{D-S}[T]
\right]^{T}.
\end{align}

The EE-TOTD problem can be formulated as follows:
\begin{small}
\begin{equation}
\begin{aligned}
\min_{\substack{
\{\boldsymbol{\alpha}_{i}^{D-U}\}_{D_i\in\mathcal D},
\{\boldsymbol{\alpha}_{i,m}^{D-S}\}_{D_i\in\mathcal D,S_m\in\mathcal S},\\
\boldsymbol{q}^{(U)},
\{\boldsymbol{f}_{i}^{D-U}\}_{D_i\in\mathcal D},
\{\boldsymbol{f}_{i,m}^{D-S}\}_{D_i\in\mathcal D,S_m\in\mathcal S}
}}
\quad & E
\end{aligned}
\label{eq:EE-TOTD}
\end{equation}
\end{small}
\vspace{-0.4cm}

\noindent
s.t.

\vspace{-0.4cm}
\begin{small}
\begin{equation*}
\begin{array}{l}
C1:\sum\limits_{t=1}^T(\alpha_i^{D - U}[t] + \sum\limits_{{S_m} \in {\cal S}} {\alpha _{i,m}^{D - S}[t]})  = 1,\forall {D_i} \in {\cal D},\\
	C2: 0 \le \alpha _i^{D - U}[t] \le 1,\forall {D_i} \in {\cal D}, t\in\mathcal{T},\\

	C3: 0 \le \alpha _{i,m}^{D - S}[t] \le 1,\forall {D_i} \in {\cal D},{S_m} \in {\cal S}, t\in\mathcal{T},\\

	C4:\sum\limits_{{D_i} \in {\cal D}} {f_i^{D - U}[t]}  \le {F^{(U)}},t\in\mathcal{T},\\

	C5:\sum\limits_{{D_i} \in {\cal D}} {f_{i,m}^{D - S}[t]}  \le {F^{(S)}}\cdot \chi_m[t],\forall {S_m} \in {\cal S},t\in\mathcal{T},\\

    C6: -R \leq x^{(U)}[t],y^{(U)}[t] \leq R, t\in\mathcal{T},\\
    
	C7:\Delta \cdot {\gamma _i^{D - U}[t]}  \ge\alpha _i^{D - U}[t] \cdot V_i^{(D)}+\sum\limits_{{S_m} \in {\cal S}} {\alpha _{i,m}^{D - S}[t]\cdot V_i^{(D)}},\\
    \quad \quad \forall {D_i} \in {\cal D},t\in\mathcal{T},\\

	C8:\Delta \cdot {\gamma _m^{U - S}[t]} \cdot \chi_m[t]  \ge \sum\limits_{{D_i} \in {\cal D}} {\alpha _{i,m}^{D - S}[t]\cdot V_i^{(D)}},\\
    \quad \quad\forall {S_m} \in {\cal S},t\in\mathcal{T},\\

	C9:\Delta  \cdot {f_i^{D - U}[t]}  \ge \alpha _i^{D - U}[t] \cdot V_i^{(D)} \cdot \eta^{(D)},\forall {D_i} \in {\cal D},t\in\mathcal{T},\\

	C10:\Delta  \cdot  {f_{i,m}^{D - S}[t]} \cdot \chi_m[t] \ge \alpha _{i,m}^{D - S}[t] \cdot V_i^{(D)} \cdot \eta^{(D)},\\
    \quad \quad\forall {D_i} \in {\cal D},{S_m} \in {\cal S},t\in\mathcal{T},\\

	C11:f_i^{D - U}[t] \ge 0,\forall {D_i} \in {\cal D},t\in\mathcal{T},\\

	C12:f_{i,m}^{D - S}[t] \ge 0,\forall {D_i} \in {\cal D},{S_m} \in {\cal S},t\in\mathcal{T}.\\

\end{array}
\end{equation*}
\end{small}

Constraint C1 guarantees that each task generated by IDs is completely processed by the UAV and LEO satellites. 
Constraints C2 and C3 define the feasible ranges of the task splitting variables, ensuring that the task portions assigned to the UAV and satellites are valid. 
Constraints C4 and C5 characterize the limited computing capabilities of the UAV and satellites, respectively. 
Specifically, the total computing resources allocated by the UAV to all tasks in each time slot cannot exceed $F^{(U)}$, and the total computing resources allocated by each satellite cannot exceed $F^{(S)}$. 
Constraint C6 ensures that the UAV stays within the allowed UAV flight region.
Constraint C7 ensures that the ID-to-UAV transmission capacity is no smaller than the total amount of task data to be offloaded in each time slot. Since the task data allocated to satellites must first be transmitted to the UAV for relaying, the ID-to-UAV transmission data includes both the task data processed at the UAV and the task data further forwarded to the satellites.
Constraint C8 guarantees that, in each time slot, the amount of task data forwarded from the UAV to satellite $S_m$ does not exceed the available UAV-to-satellite transmission capacity when $S_m$ is visible.
Constraint C9 requires that the UAV completes its assigned portion of task $D_i$ in each time slot. 
Constraint C10 ensures that if the satellite $S_m$ is available in that time slot, task $D_i$ allocated to $S_m$ must be completed within that time slot.
Finally, constraints C11 and C12 ensure the non-negativity of the computing resource allocation variables.

It is worth noting that problem \eqref{eq:EE-TOTD} is non-convex due to the coupled communication, computation, and trajectory design variables. In particular, the achievable transmission rates depend on the ID transmit power, UAV position, and beamforming vectors, while the total energy consumption is also coupled with task splitting and computing resource allocation. 
As a result, the EE-TOTD problem is difficult to solve directly. Therefore, it is decomposed into two tractable subproblems for efficient optimization.

\subsection{Subproblem: UAV Trajectory Planning}
For given $\alpha_i^{*D - U}[t],\alpha_{i,m}^{*D - S}[t],f_i^{*D - U}[t],f_{i,m}^{*D - S}[t]$, the energy consumption in (\ref{eq:total energy}) can be rewritten as
\begin{small}
	\begin{equation}
		\begin{array}{r}
			E  =\sum\limits_{t = 1}^T { {\left( {E _{cpt}^{*(U)}[t] + E _{trs}^{*(U)}[t] + E _{ppl}^{(U)}[t]} \right)}}\\
            +\sum\limits_{t = 1}^T {\sum\limits_{{D_i} \in D} {\left( {E _{trs,i}^{*(D)}[t]} \right)} }+ \sum\limits_{t = 1}^T {\sum\limits_{{S_m} \in S} {E _{cpt,m}^{*(S)}[t]} },
		\end{array}
		\label{eq:total energy_B}
	\end{equation}
\end{small}
where $E _{cpt}^{*(U)}[t],E _{trs}^{*(U)}[t],E _{trs,i}^{*(D)}[t]$, and $E _{cpt,m}^{*(S)}[t]$ are constant terms determined by the fixed transmission powers of IDs and UAV, as well as the given task-splitting ratios and computing resource allocation.
Accordingly, the EE-TOTD problem in (\ref{eq:EE-TOTD}) can be reformulated as a UAV trajectory planning subproblem below
\begin{small}
\begin{equation}
\begin{aligned}
&\quad\quad\quad\quad\quad\min \limits_{tr^{(U)}[t]} \sum\limits_{t = 1}^T {{{E_{ppl}^{(U)}[t]} }}\\
&\rm{s.t.}\quad C6,\\
&\quad\quad C^*1:\Delta \cdot {\gamma _i^{D - U}[t]}  \ge\alpha _i^{*D - U}[t] \cdot V_i^{(D)}+\\
    &\quad\quad\sum\limits_{{S_m} \in {\cal S}} {\alpha _{i,m}^{*D - S}[t]\cdot V_i^{(D)}}, t\in\mathcal{T},
\label{eq:EE-TOTD_UAV}
\end{aligned}
\end{equation}
\end{small}

\noindent
where constraint $\mathrm{C}^*1$ guarantees the feasibility of ID-to-UAV data transmission by requiring the UAV to stay sufficiently close to each served ID. In this way, the corresponding ID-to-UAV channel capacity can support the amount of task data transmitted from the ID to the UAV.
For given task splitting and computing resource allocation, the UAV trajectory subproblem still involves nonlinear propulsion energy and trajectory-dependent communication constraints. 
Therefore, it remains a non-convex optimization problem.

\subsection{Subproblem: Task Splitting and Computing Resource Allocation}
For a given UAV trajectory $tr^{*(U)}[t]$, the energy consumption in (\ref{eq:total energy}) can be rewritten as
\begin{small}
	\begin{equation}
		\begin{array}{r}
			E  =\sum\limits_{t = 1}^T { {\left( {E_{cpt}^{(U)}[t] + E _{trs}^{*(U)}[t] + E _{ppl}^{*(U)}[t]} \right)}}\\
            +\sum\limits_{t = 1}^T {\sum\limits_{{D_i} \in D} {\left( {E _{trs,i}^{*(D)}[t]} \right)} }+ \sum\limits_{t = 1}^T {\sum\limits_{{S_m} \in S} {E _{cpt,m}^{(S)}[t]} } ,
		\end{array}
		\label{eq:sub total energy}
	\end{equation}
\end{small}

\noindent
where $E _{trs}^{*(U)}[t], E _{trs,i}^{*(D)}[t]$, and $E _{ppl}^{*(U)}[t]$ are constant terms determined by the fixed transmission power of IDs and UAV, as well as the given UAV trajectory.
Accordingly, the EE-TOTD problem in (\ref{eq:EE-TOTD}) can be reformulated as
\begin{small}
\begin{equation}
\mathop {\arg }\limits_{\scriptstyle\alpha_i^{D - U}[t],\alpha_{i,m}^{D - S}[t],\atop
{f_i^{D - U}[t],f_{i,m}^{D - S}[t],\atop}} \min \left(\sum\limits_{t = 1}^T { {{E_{cpt}^{(U)}[t]} }}+ \sum\limits_{t = 1}^T {\sum\limits_{{S_m} \in S} {E _{cpt,m}^{(S)}[t]} }+E^*_c\right), \atop
\rm{s.t.} \quad C1-C5, C7-C12,
\label{eq:EE-TOTD_1}
\end{equation}
\end{small}

\noindent
where $E^*_c$ denotes the sum of all constant energy terms in (\ref{eq:sub total energy}).

According to constraints C9 and C10, the minimum value of subproblem (\ref{eq:EE-TOTD_1}) can be formulated as

\begin{small}
\begin{equation}
\begin{aligned}
E = & E^*_c+\sum\limits_{t = 1}^T \sum\limits_{D_i \in \mathcal{D}}
\rho
\left( f_{i}^{D-U}[t] \right)^2
\alpha_{i}^{D-U}[t] V_i^{(D)} \eta^{(D)}
\\
& + \sum\limits_{t = 1}^T \sum\limits_{S_m \in \mathcal{S}}
\left(
\sum\limits_{D_i \in \mathcal{D}}
\rho
\left( f_{i,m}^{D-S}[t] \right)^2
\alpha_{i,m}^{D-S}[t] V_i^{(D)} \eta^{(D)}
\right)
\\
& \geq E^*_c+\frac{\rho}{\Delta^2}\sum\limits_{t = 1}^T \sum\limits_{D_i \in \mathcal{D}}(\alpha_{i}^{D-U}[t] V_i^{(D)} \eta^{(D)})^3 \\
&+\frac{\rho}{\Delta^2}\sum\limits_{t = 1}^T \sum\limits_{S_m \in \mathcal{S}} \left(
\sum\limits_{D_i \in \mathcal{D}}(\alpha_{i,m}^{D-S}[t] V_i^{(D)} \eta^{(D)})^3\right).
\end{aligned}
\label{eq:min value}
\end{equation}
\end{small}
We use $E_{\min}$ to denote the minimum value on the right-hand side of (\ref{eq:min value}). Let $\mathcal{A}=\{U,S_1,S_2,\ldots,S_M\}$ denote the set of computing nodes in the SAGIIN, where $|\mathcal{A}|=M+1$. For notation clarity, we use $A_n \in \mathcal{A}$ to represent a generic computing node. Specifically, $A_1$ denotes the UAV, while $A_{m+1}$ denotes satellite $S_m$.
Therefore, we can rewrite the task splitting variables as
\begin{equation}
\alpha_{i,n=1}[t]=\alpha_i^{D-U}[t],
\end{equation}
and
\begin{equation}
\alpha_{i,n=m+1}[t]=\alpha_{i,m}^{D-S}[t].
\end{equation}
Accordingly, $E_{\min}$ can be rewritten as
\begin{small}
\begin{equation}
E_{\min}=E^*_c+\frac{\rho}{\Delta^2}\sum\limits_{t = 1}^T \sum\limits_{D_i \in \mathcal{D}}\sum\limits_{A_n \in \mathcal{A}}(\alpha_{i,n}[t] V_i^{(D)} \eta^{(D)})^3.
\end{equation}
\end{small}
Therefore, subproblem (\ref{eq:EE-TOTD_1}) can be reformulated into 
\begin{small}
\begin{equation}
\begin{aligned}
\min  \limits_{\alpha_{i,n}[t]} &\left(E^*_c+\frac{\rho}{\Delta^2}\sum\limits_{t = 1}^T \sum\limits_{D_i \in \mathcal{D}}\sum\limits_{A_n \in \mathcal{A}}(\alpha_{i,n}[t] V_i^{(D)} \eta^{(D)})^3\right), \\
&\rm{s.t.} \\
&\quad C'1 : \Delta \cdot {\gamma _i^{*D - U}[t]}  \geq \sum\limits_{A_n \in \mathcal{A}}\alpha_{i,n}[t] \cdot V_i^{(D)},\\
&\quad\quad\quad\forall {D_i} \in {\cal D},t\in\mathcal{T},\\
&\quad C'2:\Delta \cdot {\gamma _{m}^{*U - S}[t]} \cdot \chi_m[t]  \ge \sum\limits_{{D_i} \in {\cal D}} {\alpha _{i,n=m+1}[t]\cdot V_i^{(D)}},\\
&\quad\quad\quad\forall {S_m} \in {\cal S},t\in\mathcal{T},\\
&\quad C'3:\sum\limits_{{D_i} \in {\cal D}} \frac{\alpha_{i,n=1}[t] V_i^{(D)} \eta^{(D)}}{\Delta}\le {F^{(U)}},t\in\mathcal{T},\\
&\quad C'4:\sum\limits_{{D_i} \in {\cal D}} \frac{\alpha_{i,m+1}[t] V_i^{(D)} \eta^{(D)}}{\Delta}\le {F^{(S)}} \cdot \chi_m[t],\\
&\quad \quad \quad \forall {S_m} \in {\cal S}, t\in\mathcal{T},\\
&\quad C'5: \sum\limits_{t=1}^T\sum\limits_{A_n \in \mathcal{A}}\alpha_{i,n}[t]=1,\forall {D_i} \in {\cal D},\\
&\quad C'6:\alpha_{i,n}[t]\geq 0,\forall {D_i} \in {\cal D},{A_n} \in {\cal A}, t\in\mathcal{T},
\label{eq:EE-TOTD_2}
\end{aligned}
\end{equation}
\end{small}

\noindent
where $\gamma _i^{*D - U}[t]$ and $\gamma _{m}^{*U - S}[t]$ are constant values due to the given UAV trajectory and the fixed transmission power of IDs and UAV.

The convexity of (\ref{eq:EE-TOTD_2}) is analyzed as follows.

\textbf{Proposition 1:} The task splitting subproblem in (\ref{eq:EE-TOTD_2}) is a convex optimization problem.

\textit{Proof:}
Since $\alpha_{i,n}[t]$ denotes the task splitting ratio, it satisfies
$\alpha_{i,n}[t]\geq 0$. For notational simplicity, define
\begin{equation}
c_i = V_i^{(D)}\eta^{(D)}.
\end{equation}
Since $V_i^{(D)}>0$ and $\eta^{(D)}>0$, we have $c_i>0$. 
The objective function of (\ref{eq:EE-TOTD_2}) can then be written as
\begin{equation}
E_{\min}
=
E_c^*
+
\frac{\rho}{\Delta^2}
\sum_{t=1}^{T}
\sum_{D_i\in\mathcal D}
\sum_{A_n\in\mathcal A}
\left(c_i\alpha_{i,n}[t]\right)^3 .
\end{equation}
For each variable $\alpha_{i,n}[t]$, the second-order derivative of 
$\left(c_i\alpha_{i,n}[t]\right)^3$ is
\begin{equation}
\frac{\partial^2 \left(c_i\alpha_{i,n}[t]\right)^3}
{\partial \left(\alpha_{i,n}[t]\right)^2}
=
6c_i^3\alpha_{i,n}[t].
\end{equation}
Since $c_i>0$ and $\alpha_{i,n}[t]\geq 0$, we have
\begin{equation}
6c_i^3\alpha_{i,n}[t]\geq 0.
\end{equation}
Thus, $\left(c_i\alpha_{i,n}[t]\right)^3$ is convex over 
$\alpha_{i,n}[t]\geq 0$. 
Since a non-negative weighted sum of convex functions is still convex and $E_c^*$ is a constant, the objective function of (\ref{eq:EE-TOTD_2}) is convex.
Besides, all constraints in (\ref{eq:EE-TOTD_2}) are affine, and the feasible set is convex. 
Therefore, the task splitting subproblem in (\ref{eq:EE-TOTD_2}) is a convex optimization problem, which can be solved using standard convex optimization solvers like CVXPY.

\section{Algorithm Design}
\label{algorithm}
In this section, we present the proposed RLD algorithm for solving the formulated EE-TOTD problem. The original problem is decomposed into two subproblems. 
The UAV trajectory planning subproblem is solved by RLD method, while the task splitting and computing resource allocation subproblem is solved using CVXPY.
Based on the solutions to the two subproblems, an alternating optimization algorithm is devised to solve the formulated EE-TOTD problem.

\subsection{Overall Framework}


Given the highly time-varying nature of SAGIIN environments, an offline training and online decision-making framework is adopted to ensure practical applicability.
However, conventional noise-based exploration may generate inefficient or infeasible UAV trajectories, such as those that deviate significantly from mobile IDs, involve abrupt movements, or approach the flight boundary. 
To address these limitations, the RAG-LLM module provides context-aware trajectory guidance by considering ID mobility patterns, satellite service windows, and retrieved historical cases with low energy consumption. 
The resulting structured trajectories provide high-quality experiences for the shared replay buffer, thereby improving DDPG sample efficiency in dynamic SAGIIN environments.

\begin{figure}[t]
\centering
\resizebox{0.98\columnwidth}{!}{%
\begin{tikzpicture}[
    x=1cm,
    y=1cm,
    >=stealth,
    font=\sffamily\small,
    block/.style={
        rectangle,
        rounded corners=2pt,
        draw=black,
        fill=white,
        line width=0.8pt,
        align=center,
        minimum height=0.95cm,
        text width=3.7cm,
        inner sep=4pt
    },
    mainblock/.style={
        rectangle,
        rounded corners=2pt,
        draw=black,
        fill=white,
        line width=0.8pt,
        align=center,
        minimum height=1.0cm,
        text width=4.2cm,
        inner sep=4pt
    },
    decision/.style={
        diamond,
        aspect=2.2,
        draw=black,
        fill=white,
        line width=0.8pt,
        align=center,
        text width=2.8cm,
        inner sep=2pt
    },
    flow/.style={->, draw=black, line width=0.8pt},
    feedback/.style={->, dashed, draw=black, line width=0.8pt}
]

\node[block, text width=4.0cm, minimum height=0.85cm]
    (scenario) at (7.2,14.7)
    {Select an ID mobility scenario\\from the training set};

\node[decision] (source) at (7.2,12.5)
    {Select trajectory source\\for episode $e$};

\draw[flow] (scenario.south) -- (source.north);

\draw[flow] (source.south) -- (7.2,11.1);
\draw (3.2,11.1) -- (11.2,11.1);

\node[fill=white, inner sep=1.5pt, align=center]
    at (3.6,11.6)
    {$e<N_{\mathrm{warm}}$ or\\$u_e<p_{\mathrm{RAG}}$};

\node[fill=white, inner sep=1.5pt]
    at (10,11.6) {otherwise};

\node[block] (retrieve) at (3.2,10.0)
    {Retrieve relevant cases\\from case memory $\mathcal{M}$};

\node[block] (rag) at (3.2,8.1)
    {RAG-LLM generates a UAV trajectory};

\draw[flow] (3.2,11.1) -- (retrieve.north);
\draw[flow] (retrieve.south) -- (rag.north);

\node[block] (actor) at (11.2,10.0)
    {DDPG actor generates\\a UAV trajectory};

\draw[flow] (11.2,11.1) -- (actor.north);

\coordinate (merge) at (7.2,6.75);
\fill[black] (merge) circle (1.5pt);

\draw[flow] (rag.south) -- (merge);
\draw[flow] (actor.south) -- (merge);

\node[mainblock, text width=4.4cm, minimum height=1.15cm]
    (cvx) at (7.2,5.0)
    {CVXPY solves the task-splitting and computing-resource optimization subproblem};

\draw[flow] (merge) -- (cvx.north);

\node[mainblock, text width=4.2cm, minimum height=0.95cm]
    (rewards) at (7.2,2.5)
    {Obtain energies and assign rewards};

\draw[flow] (cvx.south) -- (rewards.north);

\node[block, text width=3.2cm, minimum height=1.05cm]
    (memory) at (2.0,2.5)
    {Case memory $\mathcal{M}$\\
     Store evaluated trajectory and energy};

\draw[flow] (rewards.west) -- (memory.east);

\node[mainblock, text width=4.0cm, minimum height=0.95cm]
    (buffer) at (7.2,0.45)
    { Store in replay buffer $\mathcal{B}$
     };

\draw[flow] (rewards.south) -- (buffer.north);

\node[block, text width=3.2cm, minimum height=1.05cm]
    (update) at (11.5,0.45)
    {DDPG updates actor and critic networks};

\draw[flow] (buffer.east) -- (update.west);

\draw[feedback]
    (memory.west) -- (0.1,2.5) -- (0.1,10.0) -- (retrieve.west);

\draw[feedback]
    (update.east) -- (13.7,0.45) -- (13.7,10.0) -- (actor.east);

\node[rotate=90, fill=white, inner sep=2pt]
    at (-0.1,6.1) {retrieve in later episodes};

\node[rotate=90, fill=white, inner sep=2pt]
    at (13.98,6.1) {updated policy};

\end{tikzpicture}
}
\caption{The episode-level RLD algorithm framework.}
\label{fig:rld_overview}
\end{figure}
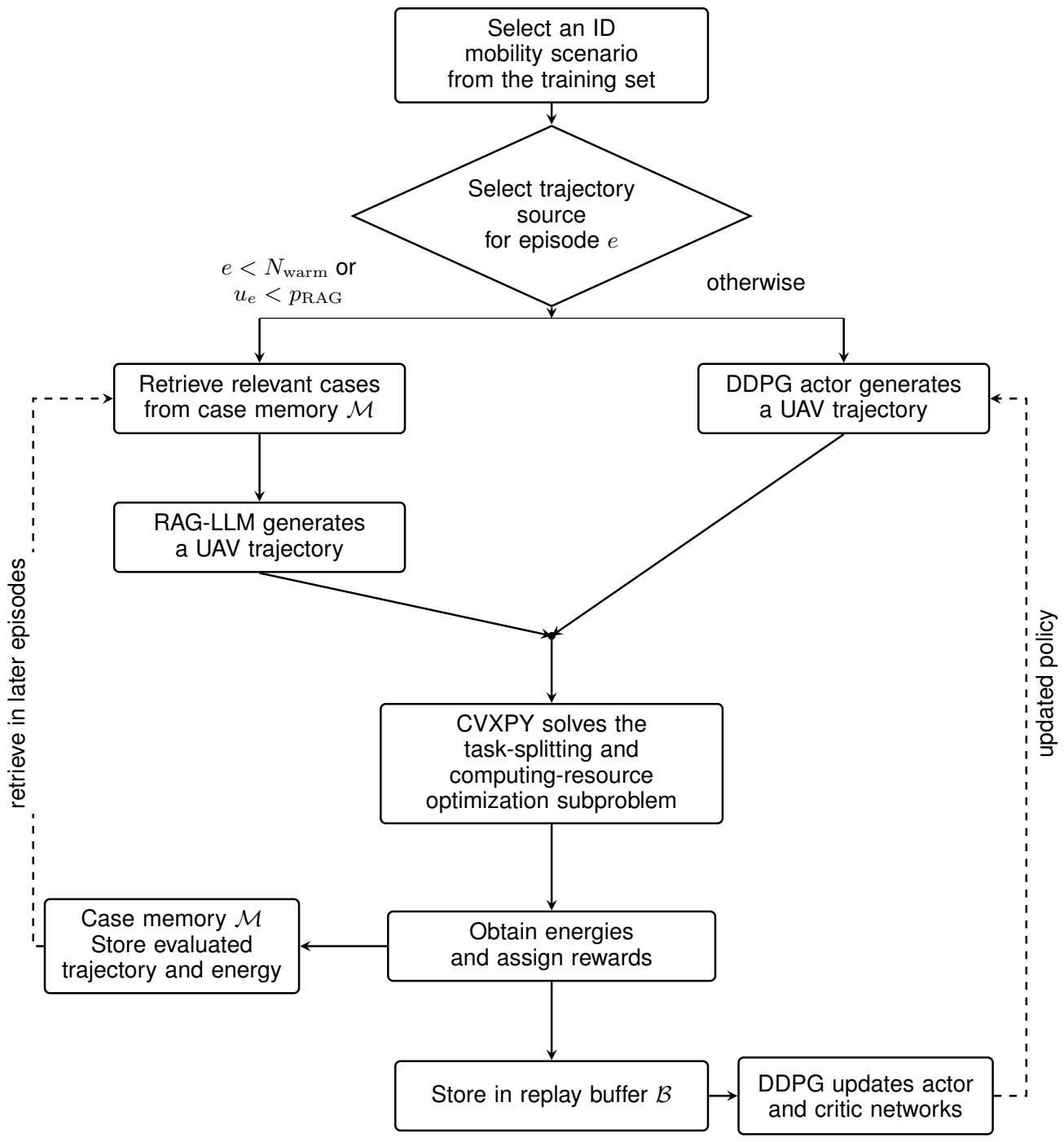

The RLD algorithm comprises three interconnected modules: a case-based RAG-LLM module, a DDPG learning module, and a CVXPY-based resource optimization module. 
The framework of RLD is shown in Fig. \ref{fig:rld_overview}.
At the beginning of each training episode, an ID mobility scenario is selected from the predefined training scenario set. 
The trajectory source is then selected according to
\begin{equation}
\mathbf Q_e^{(U)}=
\begin{cases}
\mathbf Q_{e,\mathrm{RAG-LLM}}^{(U)},
& e<N_{\mathrm{warm}}\ \mathrm{or}\ u_e<p_{\mathrm{RAG}},\\
\mathbf Q_{e,\mathrm{DDPG}}^{(U)},
& \mathrm{otherwise},
\end{cases}
\label{eq:episode_source_selection}
\end{equation}
where $u_e\sim\mathcal U(0,1)$, $N_{\mathrm{warm}}$ is the number of warm-up episodes, and $p_{\mathrm{RAG}}$ is the probability of selecting the RAG-LLM branch after warm-up.

Regardless of the selected trajectory source, a complete UAV trajectory is first generated and then evaluated by the CVXPY-based module, which solves the entire period task splitting and computing resource optimization problem and returns the resulting slot-wise energy consumption. These energy values are converted into post-evaluation rewards and assigned to the corresponding slot-wise state-action transitions. The transitions are then inserted into the shared replay buffer $\mathcal{B}$, from which the DDPG module samples experience to update its actor and critic networks. The evaluated trajectory and its energy performance are also retained in the case memory for retrieval in subsequent RAG-LLM episodes. 

\subsection{DDPG-Based UAV Trajectory Planning}
\label{DDPG}
The state of the DDPG agent in time slot $t$ is defined as
\begin{small}
\begin{equation}
\begin{aligned}
s[t]=\big[&
\widetilde{\mathbf Q}^{(D)}[t],
\widetilde{\mathbf q}^{(U)}[t-1],
\widetilde{\boldsymbol{\nu}}^{(U)}[t-1],
{\boldsymbol{\chi}}[t],
t/T
\big],\\
&\widetilde{\mathbf Q}^{(D)}[t]
=
\{\widetilde{\mathbf q}^{(D)}_1[t],\ldots,
\widetilde{\mathbf q}^{(D)}_I[t]\},\\
&{\boldsymbol{\chi}}[t]
=
\{\chi_1[t],\ldots,\chi_M[t]\},
\end{aligned}
\label{state}
\end{equation}
\end{small}
where $\mathbf q_i^{(D)}[t]=[X_i^{(D)}[t], Y_i^{(D)}[t]]$ denotes the horizontal position of ID $D_i$, $\chi_m[t]$ represents the service availability of satellite $S_m$, $\mathbf q^{(U)}[t-1]=[x^{(U)}[t-1],y^{(U)}[t-1]]$ denotes the horizontal position of the UAV in time slot $t-1$, and $\boldsymbol{\nu}^{(U)}[t-1]$ denotes the UAV velocity in time slot $t-1$. The normalized time index $t/T$ distinguishes different stages within a scheduling period, and $\widetilde{(\cdot)}$ denotes a normalized variable.

The action is defined as the horizontal displacement of the UAV in time slot $t$:
\begin{equation}
{\bf a}[t]
=
\Delta {\bf q}^{(U)}[t]
=
\left[
\Delta x^{(U)}[t],
\Delta y^{(U)}[t]
\right].
\label{action}
\end{equation}

Given the UAV position in time slot $t-1$, the UAV position in time slot $t$ is updated by
\begin{equation}
{\bf q}^{(U)}[t]
=
{\bf q}^{(U)}[t-1]
+
\Delta {\bf q}^{(U)}[t],
\label{uav_position_update}
\end{equation}
where ${\bf q}^{(U)}[t]=[x^{(U)}[t],y^{(U)}[t]]$ denotes the horizontal position of the UAV in time slot $t$. Accordingly, the UAV velocity in time slot $t$ is given by
\begin{equation}
\boldsymbol{\nu}^{(U)}[t]
=
\frac{
\Delta {\bf q}^{(U)}[t]
}{\Delta}.
\label{uav_velocity_from_action}
\end{equation}

The displacement action sequence over $T$ time slots determines the complete UAV trajectory:
\begin{equation}
{\bf Q}^{(U)}
=
\{
{tr}^{(U)}[1],
{tr}^{(U)}[2],
\ldots,
{tr}^{(U)}[T]
\},
\label{uav_complete_trajectory}
\end{equation}
where each ${tr}^{(U)}[t]=[x^{(U)}[t],y^{(U)}[t],Z^{(U)}]$ is obtained from the recursive position update in \eqref{uav_position_update}.

After the RAG-assisted exploration stage, the UAV trajectory is generated by the DDPG actor network. Let $\pi_{\boldsymbol{\theta}}(\cdot)$ denote the actor network and $Q_{\boldsymbol{\omega}}(\cdot)$ denote the critic network, where $\boldsymbol{\theta}$ and $\boldsymbol{\omega}$ are the corresponding network parameters. The actor network maps the state ${\bf s}[t]$ to the UAV action ${\bf a}[t]$. To encourage exploration, a decaying Gaussian noise term is added to the actor network output:
\begin{equation}
{\bf a}[t]
=
\pi_{\boldsymbol{\theta}}({\bf s}[t])
+
{\bf n}[t],
\end{equation}
where ${\bf n}[t]$ is the exploration noise. The obtained action is further projected into the feasible UAV flight region.

During each episode, the complete UAV trajectory $\mathbf{Q}^{(U)}$ is first generated by the trajectory exploration module. 
Given $\mathbf{Q}^{(U)}$, the environment computes the communication rates over all time slots and then invokes the CVXPY-based optimization module to solve the full-period task splitting problem and obtain the slot-wise energy sequence ${E^\star[t]}_{t=1}^{T}$.
Then, the post-evaluation reward for each time slot is assigned as
\begin{equation}
r[t]
=
-
\frac{E^\star[t]}{\Omega_E},
\quad t=1,\ldots,T.
\label{eq:post_evaluation_reward}
\end{equation}
Accordingly, the episode-level reward is given by
\begin{equation}
R^{(e)}
=
\sum_{t=1}^{T}r[t]
=
-\frac{1}{\Omega_E}
\sum_{t=1}^{T}E^{\star}[t].
\label{episode_reward}
\end{equation}
Thus, minimizing the total energy consumption is equivalent to maximizing the accumulated episode reward.
Subsequently, the generated trajectory is transformed into slot-wise state-action transitions. The corresponding post-evaluation rewards are assigned to these transitions, and the completed transitions $(s[t],a[t],r[t],s[t+1])$ are stored in the replay buffer $\mathcal{B}$.

The critic network is updated by minimizing the following loss function:
\begin{equation}
\mathcal{L}(\boldsymbol{\omega})
=
\frac{1}{|\mathcal{B}_s|}
\sum_{j\in\mathcal{B}_s}
\left(
Q_{\boldsymbol{\omega}}({\bf s}_j,{\bf a}_j)
-
y_j
\right)^2,
\label{eq:critic_loss}
\end{equation}
where $\mathcal{B}_s$ denotes a mini-batch sampled from the replay buffer $\mathcal{B}$, and
\begin{equation}
y_j
=
r_j
+
\Gamma(1-d_j)
Q_{\boldsymbol{\omega}^{-}}
\left(
{\bf s}_{j+1},
\pi_{\boldsymbol{\theta}^{-}}({\bf s}_{j+1})
\right).
\label{eq:target_q}
\end{equation}
Here, $\Gamma$ is the discount factor, $d_j$ is the terminal indicator, and $\boldsymbol{\theta}^{-}$ and $\boldsymbol{\omega}^{-}$ are the parameters of the target actor and target critic networks, respectively.

The actor network is updated by maximizing the expected critic value, which is equivalent to minimizing
\begin{equation}
\mathcal{L}(\boldsymbol{\theta})
=
-
\frac{1}{|\mathcal{B}_s|}
\sum_{j\in\mathcal{B}_s}
Q_{\boldsymbol{\omega}}
\left(
{\bf s}_j,
\pi_{\boldsymbol{\theta}}({\bf s}_j)
\right).
\end{equation}
The target networks are updated by
\begin{equation}
\boldsymbol{\theta}^{-}
\leftarrow
\tau\boldsymbol{\theta}
+(1-\tau)\boldsymbol{\theta}^{-},
\end{equation}
\begin{equation}
\boldsymbol{\omega}^{-}
\leftarrow
\tau\boldsymbol{\omega}
+(1-\tau)\boldsymbol{\omega}^{-},
\end{equation}
where $\tau$ is the soft update coefficient.

\subsection{Case-Based RAG-LLM Exploration for DDPG}
\subsubsection{Case Memory Construction and Retrieval}

Conventional DDPG typically relies on random noise for exploration, which may lead to unstable training and low exploration efficiency, particularly during the early training stages in high-dimensional continuous action spaces \cite{DDPG_limit}. 
To improve the quality of early-stage exploration, we introduce a case-based RAG-LLM module to generate informative UAV trajectories.

The RAG module maintains an experience memory, denoted by $\mathcal{M}$. Each historical record in $\mathcal{M}$ is written as
\begin{equation}
    \mathcal{R}_e =
    \left(
    \boldsymbol{\xi}_e,
    \mathbf{Q}_e^{(U)},
    E_e,
    \{E_e[t]\}|_{t=1}^{T}
    \right),
\end{equation}
where $\boldsymbol{\xi}_e$ is the feature vector of the network state in episode $e$, ${\bf Q}^{(U)}_e$ is the generated UAV trajectory, $E_e$ is the total energy consumption, and $E_e[t]$ is the energy consumption in time slot $t$.

For the current episode, the feature vector $\boldsymbol{\xi}$ is constructed according to the ID mobility pattern and the initial UAV position. Specifically, for each time slot, the feature vector $\boldsymbol{\xi}$ can be expressed as
\begin{small}
\begin{equation}
\boldsymbol{\xi}=
\left[
\phi^T[1],\phi^T[2],\dots,\phi^T[T]
\right]^T,
\end{equation}   
\end{small}

\begin{small}
\begin{equation}
\boldsymbol{\phi[t]}=
\left[
\bar{x}^{(D)}[t],\bar{y}^{(D)}[t],\sigma^{(D)}[t],d^{(D,U)}[t],
x^{(U)}[0],y^{(U)}[0], \boldsymbol{\chi}[t]
\right],
\end{equation}   
\end{small}
\noindent
where $(\bar{x}^{(D)}[t],\bar{y}^{(D)}[t])$ denotes the normalized centroid of all IDs in time slot $t$, $\sigma^{(D)}[t]$ denotes the normalized spatial spread of IDs, and $d^{(D,U)}[t]$ denotes the normalized average distance between the IDs and the initial UAV position.
Based on $\boldsymbol{\xi}$, the RAG module retrieves the most valuable historical records from $\mathcal{M}$, based on its relevance and utility \cite{reward_utility}.
The retrieval score of record $\mathcal{R}_e$ is defined as
\begin{equation}
\mathcal{S}(\boldsymbol{\xi},\boldsymbol{\xi}_e)
=
\left\|
\boldsymbol{\xi}-\boldsymbol{\xi}_e
\right\|_2
+
\lambda_E
\frac{E_e-E_{\min}}{E_{\max}-E_{\min}+\epsilon}.
\label{eq:retrieval_score}
\end{equation}
where $E_{\min}$ and $E_{\max}$ are the minimum and maximum energy consumption values in $\mathcal{M}$, respectively, $\lambda_E$ is a weight factor, and $\epsilon$ is a small positive constant.
The first term measures the feature distance between the current scenario and the $e$-th historical scenario. A smaller distance indicates a higher scenario similarity. The second term measures the normalized energy gap between the $e$-th historical case and the best case stored in $\mathcal{M}$. A smaller energy gap indicates a better historical solution quality. Therefore, $\lambda_E$ is used to balance the importance of historical energy performance and scenario similarity during case retrieval.

The top-$K$ retrieved records are inserted into the LLM prompt together with the current ID positions, UAV flight region, initial UAV position, and physical trajectory planning principles. The LLM is required to generate a candidate UAV trajectory:
\begin{equation}
{\bf Q}_{\rm LLM}^{(U)}
=
\{[x^{(U)}[1],y^{(U)}[1]],\ldots,[x^{(U)}[T],y^{(U)}[T]]\}.
\end{equation}

\subsubsection{Prompt Design}

After retrieving the top-$K$ historical cases, the retrieved 
information is incorporated into a structured prompt to guide the 
LLM in generating an energy-efficient UAV trajectory. The prompt 
contains four main components: the optimization objective, physical 
trajectory design principles, retrieved-case guidance, and the 
required output format. 
Specifically, the generated trajectory should follow three principles. First, the UAV should move closer to the ID cluster to guarantee the ID-to-UAV transmission rate. 
Second, unnecessary long-distance movement should be avoided to reduce propulsion energy consumption. 
Third, the UAV should remain within the feasible flight region. 
After the LLM generates the trajectory, a projection operation is used to ensure C6 and C*1.
If the LLM output is invalid, a Centroid fallback trajectory is generated.

The main prompt to generate the trajectory is shown below:

\begin{quote}
\small
\ttfamily
You design a smooth UAV trajectory for an energy-minimizing mobile 
edge computing system. Follow three physical principles:

1) Stay close to the mobile-device cluster in each time slot so that 
the ID-to-UAV transmission rate remains high.

2) Avoid long-distance or high-speed movement between consecutive 
time slots to reduce rotary-wing UAV propulsion energy consumption.

3) Keep every trajectory point inside the square flight region centered at the origin with half-side length $R$, i.e., 
$-R\leq x,y\leq R$.

Lower-energy retrieved cases should be treated as better references.

Generate $T$ two-dimensional UAV positions in the following 
JSON-array format:

[[x1,y1],[x2,y2],...,[xT,yT]]

Only return the JSON array without any additional explanation or text.
\end{quote}
In the implementation, $R$ and $T$ in the prompt are replaced by their corresponding values in the current environment.

The retrieved cases are organized in the prompt using the following template:
\begin{quote}
\small
\ttfamily
Retrieved Case $k$:\\
Reference trajectory: ${\bf Q}^{(U)}_k$\\
Total energy consumption: $E_k$
\end{quote}

\section{Simulation}
\begin{figure*}[t]
    \centering
    \includegraphics[width=1\linewidth]{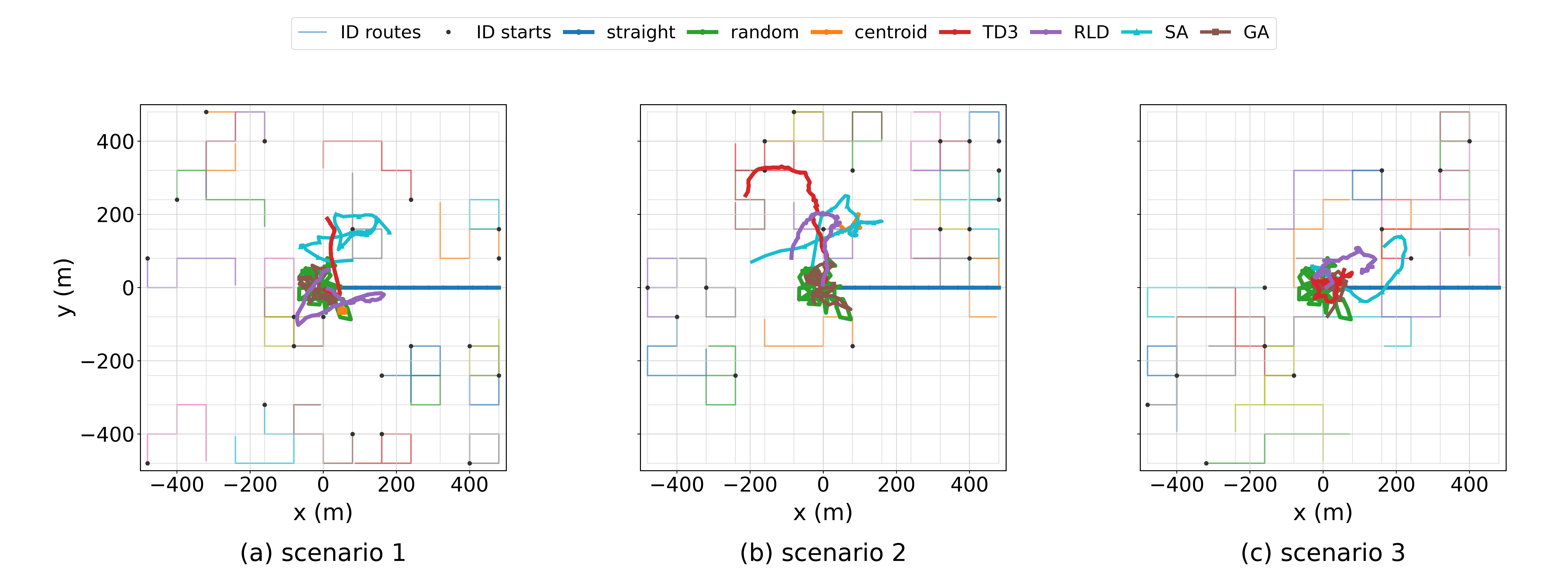}
    \caption{The UAV trajectories generated by different algorithms under the three testing scenarios.}
    \label{fig:scenario}
\end{figure*}

This section evaluates the performance of the proposed RLD algorithm for solving the optimization problem in \eqref{eq:EE-TOTD}. 
In the simulations, GPT-5.0 is employed as the agent of the LLM-based optimization module. 
To capture the mobility characteristics of IDs in suburban scenarios, we construct a street-constrained mobility model within a 1000 m $\times$ 1000 m square area centered at the origin. The street network consists of an orthogonal grid of straight roads. Each ID moves along the roads at a constant speed of 8 m/s and changes direction at intersections. In Fig. \ref{fig:scenario},\ref{fig:ID},\ref{weight_factor}, the gray lines represent the street layout, while the colored curves illustrate the paths of different IDs.
In all training and testing simulations, the service windows of the LEO satellites follow the setting illustrated in Fig. \ref{sate}. 
The remaining simulation parameters are summarized in Table \ref{simulation_parameters}.

\begin{table}[h]
\centering
\caption{Simulation Parameters}
\label{simulation_parameters}
\scriptsize
\renewcommand{\arraystretch}{1.08}
\setlength{\tabcolsep}{2pt}

\begin{tabularx}{\columnwidth}
{
>{\raggedright\arraybackslash}p{0.18\columnwidth}
>{\raggedright\arraybackslash}X
>{\centering\arraybackslash}p{0.22\columnwidth}
>{\centering\arraybackslash}p{0.17\columnwidth}
}
\hline
\textbf{Category} &
\textbf{Parameter} &
\textbf{Symbol} &
\textbf{Value} \\
\hline

\multirow{9}{*}{\makecell[l]{Network\\topology}}
& Number of IDs & $I$ & $20$--$50$ \\
& Number of LEO satellites & $M$ & $4$ \\
& Number of UAVs & -- & $1$ \\
& Number of time slots & $T$ & $60$ \\
& Slot duration & $\Delta$ & $1$ s \\
& UAV altitude & $Z^{(U)}$ & $50$ m \\
& UAV flight region & $R$ & $500$ m \\
& UAV initial position & $\mathbf{tr}^{(U)}[0]$ &
$\left[0,0,Z^{(U)}\right]$ \\
\hline


\multirow{2}{*}{\makecell[l]{Task\\model}}
& Task input data size & $V_i^{(D)}$ & $60$ MB \\
& Required CPU cycles per bit & $\eta^{(D)}$ & $100$ cycles/bit \\
\hline

\multirow{5}{*}{\makecell[l]{ID-to-UAV\\link}}
& ID transmit power & $p^{(D)}$ & $0.2$ W \\
& ID-to-UAV bandwidth & $B$ & $2$ MHz \\
& AWGN power & $\sigma_{\rm SC}^{2}$ & $-64$ dBm \\
& Reference channel gain & $\ell_{\rm ref,UAV}$ & $-20$ dB \\
& Reference distance & $\Omega_{\rm ref}$ & $1$ m \\
\hline

\multirow{9}{*}{\makecell[l]{UAV-to-\\satellite link}}
& Bandwidth & $W$ & $100$ MHz \\
& Carrier frequency & $f_c$ & $14$ GHz \\
& UAV-to-satellite distance & $d^{U-S}$ & $550$ km \\
& UAV antenna gain & $G_U$ & $24.5$ dBi \\
& Satellite antenna gain & $G_S$ & $43$ dBi \\
& UAV transmit power & $p^{U-S}$ & $0.1$ W \\
& AWGN power & $\sigma_{\rm US}^{2}$ & $-56$ dBm \\
& Rician factor & $K$ & $10$ dB \\
\hline

\multirow{2}{*}{\makecell[l]{Computing\\resource}}
& UAV computing capacity & $F^{(U)}$ & $20$ GHz \\
& Satellite computing capacity & $F^{(S)}$ & $5$ GHz \\
\hline

\multirow{8}{*}{\makecell[l]{UAV energy\\model\cite{UAV_energy}}}
& Blade profile power in hovering 
& $P_0$ & $79.86$ W \\
& Induced power in hovering 
& $P_i$ & $88.63$ W \\
& Tip speed of the rotor blade 
& $U_{\rm tip}$ & $120$ m/s \\
& Mean rotor-induced velocity in hovering 
& $v_0$ & $4.03$ m/s \\
& Fuselage drag ratio 
& $d_0$ & $0.6$ \\
& Air density 
& $\rho_{\rm air}$ & $1.225$ kg/m$^3$ \\
& Rotor solidity 
& $s$ & $0.05$ \\
& Rotor disc area 
& $A$ & $0.503$ m$^2$ \\
\hline

\end{tabularx}
\end{table}





\begin{figure}[t]
    \centering
    \includegraphics[width=1\linewidth]{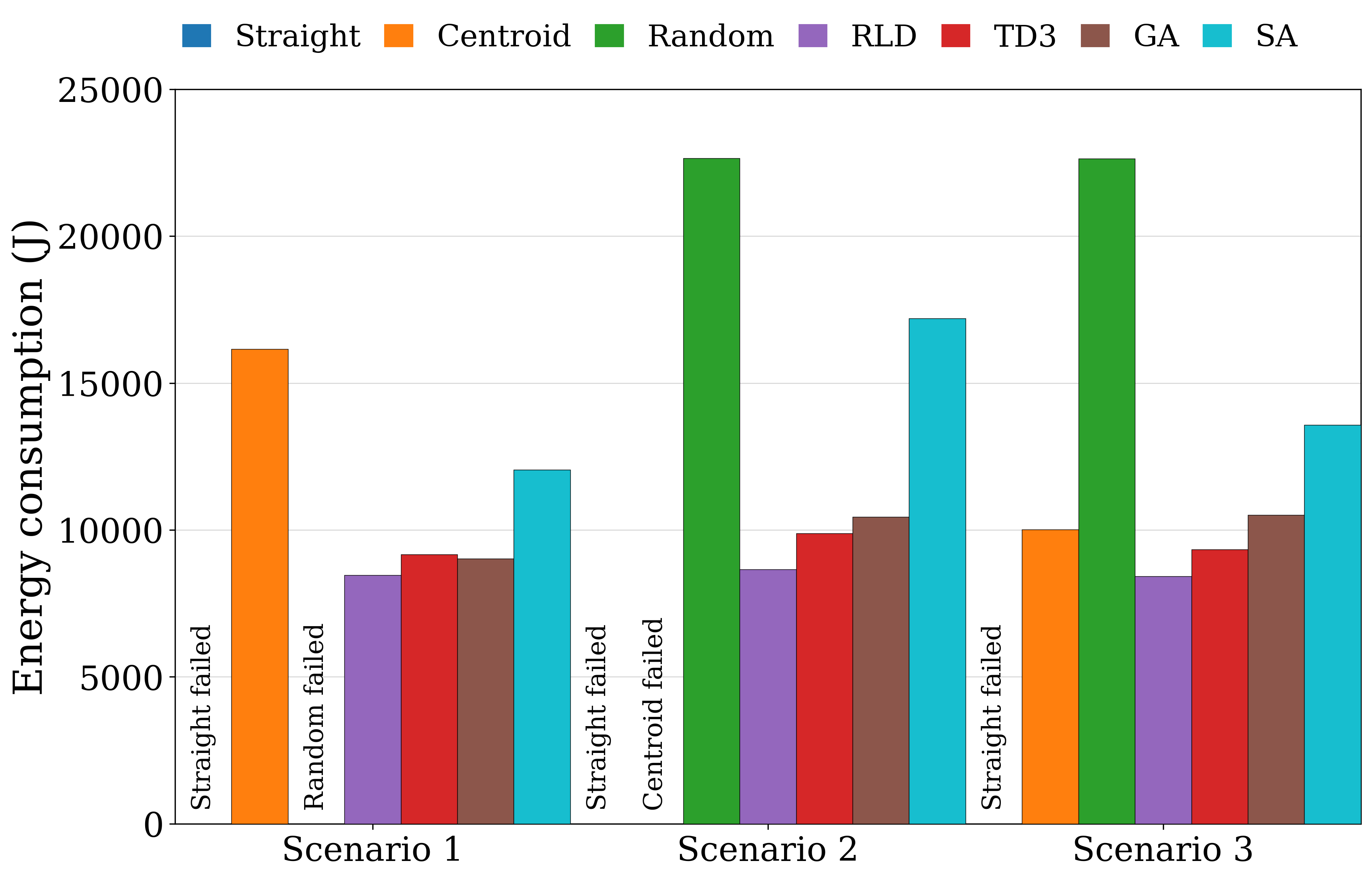}
    \caption{Energy consumption under three testing scenarios.}
    \label{scenario_energy}
\end{figure}

\begin{figure}
    \centering
    \includegraphics[width=1\linewidth]{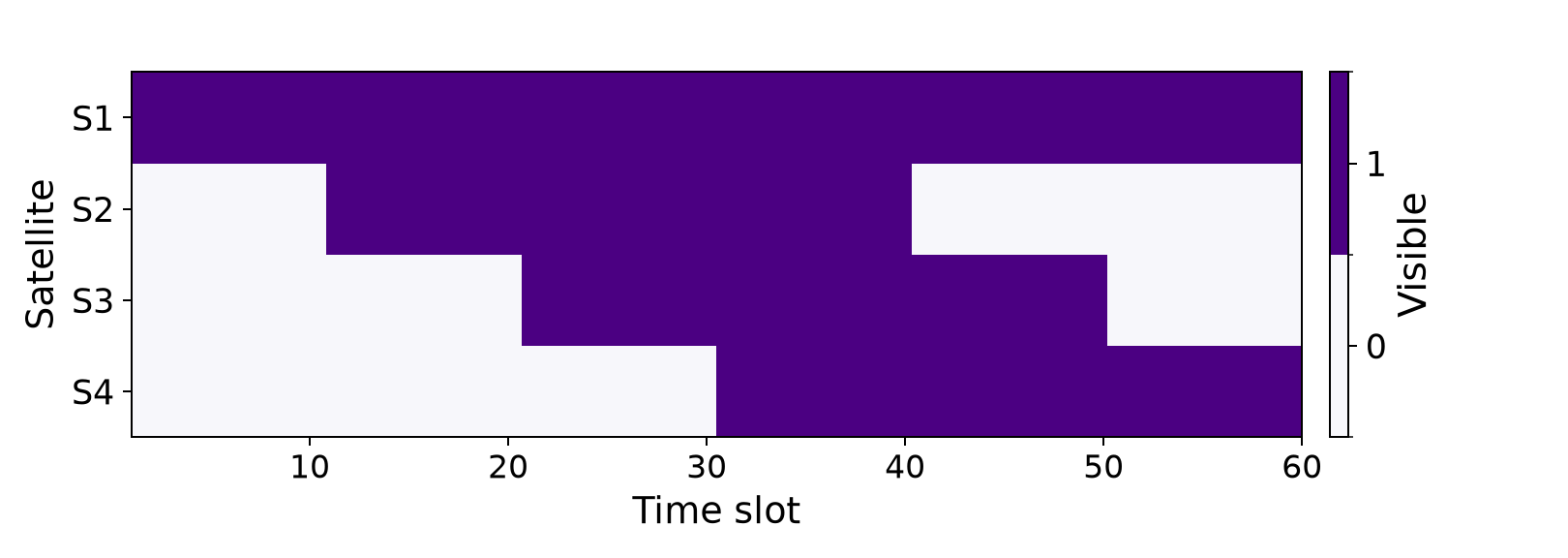}
    \caption{The service windows $\chi_m[t]$ of the LEO satellites.}
    \label{sate}
\end{figure}

During training, the RLD and DDPG models are trained on 20 different sets of ID paths generated under the same street model.  
For testing, three additional ID mobility scenarios with different spatial distributions are considered, none of which are used during training. In Scenario 1, the IDs are approximately uniformly distributed across the service area. In Scenario 2, the IDs are distributed over most of the area, but with relatively few IDs in the lower-right region. In Scenario 3, the IDs are mainly concentrated in the lower-left and upper-right regions, forming a diagonal distribution pattern.
We adopt six benchmark methods for comparison, including the straight trajectory, random trajectory, centroid trajectory, simulated annealing (SA), genetic algorithm (GA), and Twin Delayed Deep Deterministic Policy Gradient (TD3) \cite{TD3}.
TD3 adopts the twin-critic architecture, uses the minimum of the two target-critic estimates to reduce overestimation bias, and applies target-policy smoothing and delayed policy updates to improve training stability.
For the straight trajectory, the UAV flies eastward at a constant speed of 8 m/s. 
For the random trajectory, the UAV moves at a constant speed of 30 m/s, with its heading angle independently sampled from a uniform distribution over $[0,2\pi)$ in each time slot.
In the centroid trajectory, the UAV moves toward the centroid of all IDs in each time slot.

\begin{figure*}[t]
    \centering
    \includegraphics[width=1\linewidth]{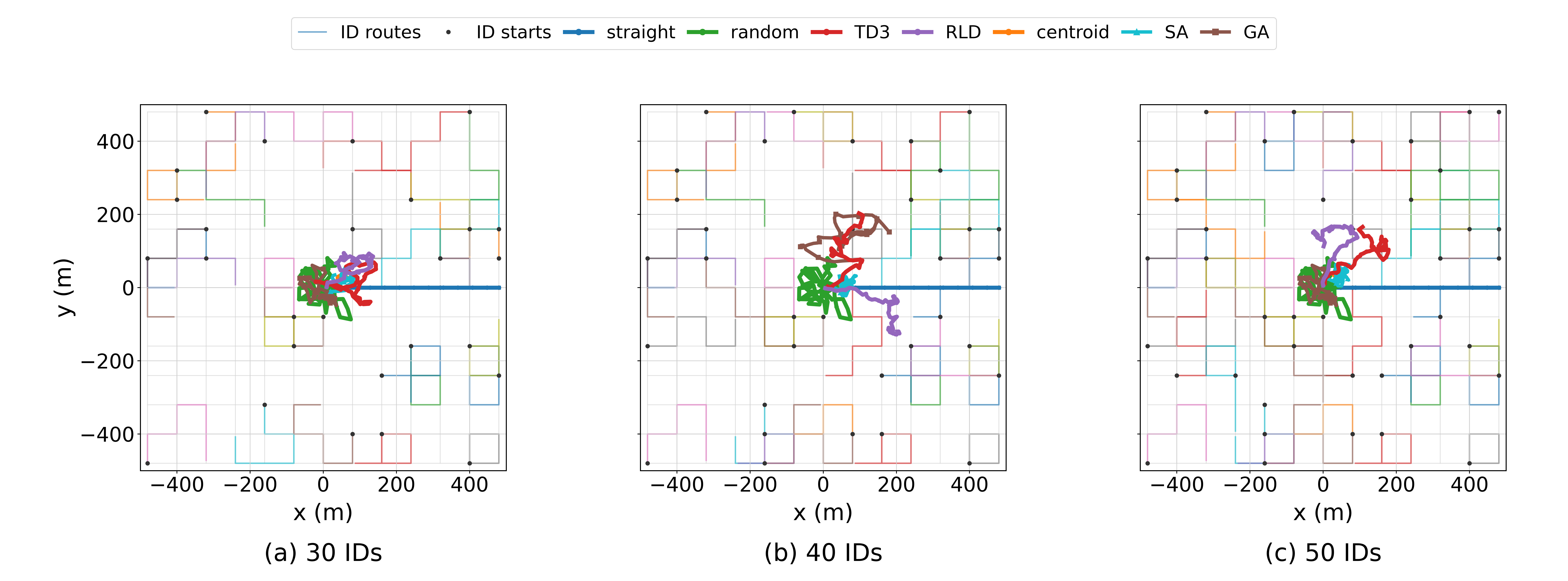}
    \caption{The UAV trajectories generated by different algorithms with different ID numbers.}
    \label{fig:ID}
\end{figure*}

\begin{figure}[t]
    \centering
    \includegraphics[width=1\linewidth]{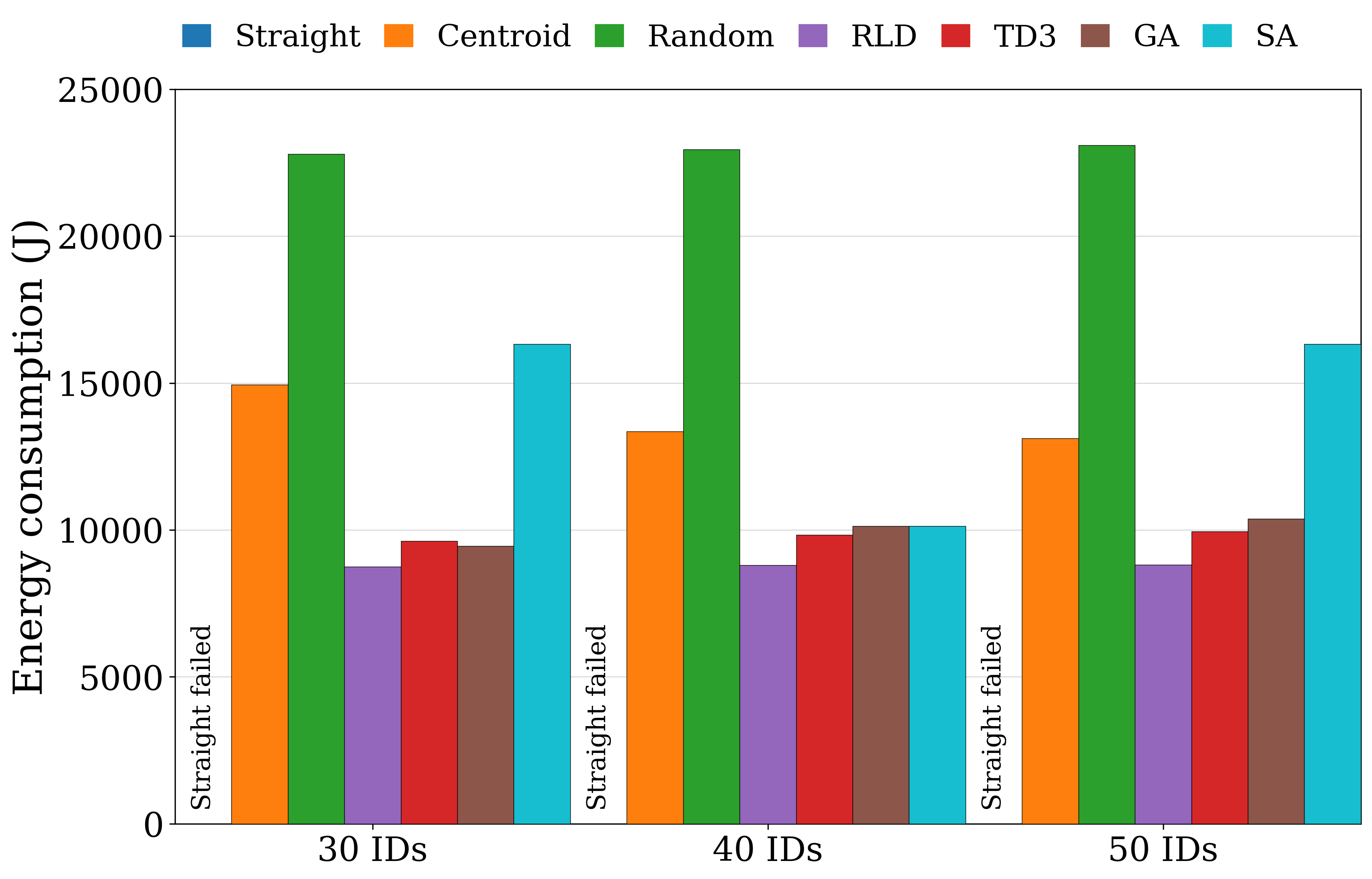}
    \caption{Energy consumption with different numbers of IDs.}
    \label{id_number_energy}
\end{figure}

Fig. \ref{fig:scenario} illustrates the UAV trajectories generated by different algorithms under the three testing scenarios, respectively. 
Fig. \ref{scenario_energy} reports the corresponding total energy consumption calculated according to \eqref{eq:total energy}. 
In the figure, \textit{Failed} indicates that the generated trajectory cannot enable the UAV to collect all task data generated by the IDs, and thus the corresponding service is regarded as failed.

The results show that both the proposed RLD algorithm and the TD3 algorithm can complete the transmission and processing of all task data in the three testing scenarios, while achieving significantly lower energy consumption than the other benchmark trajectories.
In contrast, the straight trajectory fails in all scenarios, indicating that a fixed flying direction cannot adapt to the mobility of IDs. The centroid trajectory and the random trajectory can only complete the task processing in part of the scenarios.
Although the centroid trajectory succeeds in Scenario 1, it leads to high energy consumption because the UAV needs to frequently change its direction and speed when tracking the centroid of IDs. 
Such frequent velocity variations significantly increase the UAV propulsion energy. 
The random trajectory succeeds only in Scenario 2, mainly because the IDs are relatively concentrated in this scenario.
Compared with the TD3 algorithm, the proposed RLD algorithm reduces the total energy consumption by approximately 7\%-12\%. 
This improvement is mainly because conventional noise-based exploration cannot provide prior information on promising movement directions, which may lead to inefficient UAV movements during trajectory generation. 
By retrieving historical cases with similar mobility patterns, the RAG-assisted LLM module provides informative exploration guidance for the DDPG agent, thereby reducing unnecessary UAV movement and propulsion energy consumption. 
RLD also generates more compact UAV trajectories that remain concentrated around the regions where IDs require service.
In Scenario 1, where the IDs are approximately uniformly distributed, the UAV moves back and forth to maintain coverage of as many IDs as possible. In Scenario 2, the UAV mainly moves toward the upper-left region, since no IDs require service in the lower-right region. In Scenario 3, the UAV moves close to the diagonal to maintain coverage of the IDs concentrated along this region. 
These results demonstrate that the proposed RLD algorithm can effectively improve trajectory exploration and generate more energy-efficient UAV trajectories under dynamic ID mobility.

To further evaluate the scalability of the proposed RLD algorithm under different network scales, we test its performance with different numbers of IDs. 
For each ID-number setting, the model is first trained using 20 different ID paths and then tested using an additional path that is different from the training paths. It is worth noting that increasing the number of IDs also increases the number of generated tasks proportionally. 
Under the original bandwidth and computing capacity settings, the system cannot support task transmission and processing even under ideal trajectory planning. 
Therefore, the original channel bandwidth is scaled by factors of 1.25, 1.5, and 1.75, while the computing frequencies of the UAV and satellites are scaled by factors of 1.5, 2.0, and 2.5, respectively. The ID paths and UAV trajectories are shown in Fig. \ref{fig:ID}, and the corresponding performance of different algorithms is reported in Fig. \ref{id_number_energy}.
Similar to the case with 20 IDs, the straight trajectory fails to satisfy the data transmission and processing requirements in all tested settings. 
The centroid trajectory can complete the tasks under different ID-number settings.
This is mainly because the bandwidth and computing capacity constraints are relaxed, and the increase in the number of IDs makes the movement of the centroid less drastic, thereby reducing unnecessary UAV movement. 
Similarly, the random UAV trajectory also benefits from the increased bandwidth and computing resources and can complete the tasks in these settings. 
Compared with the TD3 algorithm, the proposed RLD algorithm reduces the total energy consumption by approximately 9\%--11\%, which further demonstrates its effectiveness under different ID densities.

\begin{table}[h]
\centering
\caption{Online Decision Latency and Optimization Calls}
\label{tab:decision_latency}
\scriptsize
\renewcommand{\arraystretch}{1.08}
\setlength{\tabcolsep}{6pt}
\resizebox{\columnwidth}{!}{
\begin{tabular}{ccc}
\hline
\textbf{Algorithm}
& \textbf{Online Decision Latency (s)}
& \textbf{Optimization Calls}
\\
\hline
TD3 & 0.3477 & 1 \\
RLD       & 0.3547 & 1 \\
SA        & 23.09  & 81 \\
GA        & 103.05 & 364 \\
\hline
\end{tabular}
}
\end{table}

Table \ref{tab:decision_latency} reports the online decision latency and the number of optimization calls for different algorithms, where the number of optimization calls indicates how many times the corresponding optimization module is invoked to generate a single online decision.
The proposed RLD algorithm requires only 0.3547 s for each online decision and invokes the algorithm module once, which is comparable to the TD3 benchmark with a latency of 0.3477 s. 
This is because the proposed RLD algorithm follows an offline training and online decision-making paradigm.
The additional latency introduced by RLD is only approximately 2.01\%, indicating that the proposed mechanism incurs negligible online computational overhead. 
By contrast, SA and GA require 23.09 s and 103.05 s, respectively. 
Consequently, RLD reduces the decision latency by approximately 98.46\% relative to SA and 99.66\% relative to GA. 
Moreover, SA and GA invoke the algorithm modules 81 and 364 times, respectively, whereas RLD requires only one call. 
These results confirm that RLD avoids the repeated optimization operations required by conventional search-based algorithms, making it more suitable for delay-sensitive real-time task-offloading scenarios.

\begin{figure}
    \centering
    \includegraphics[width=1\linewidth]{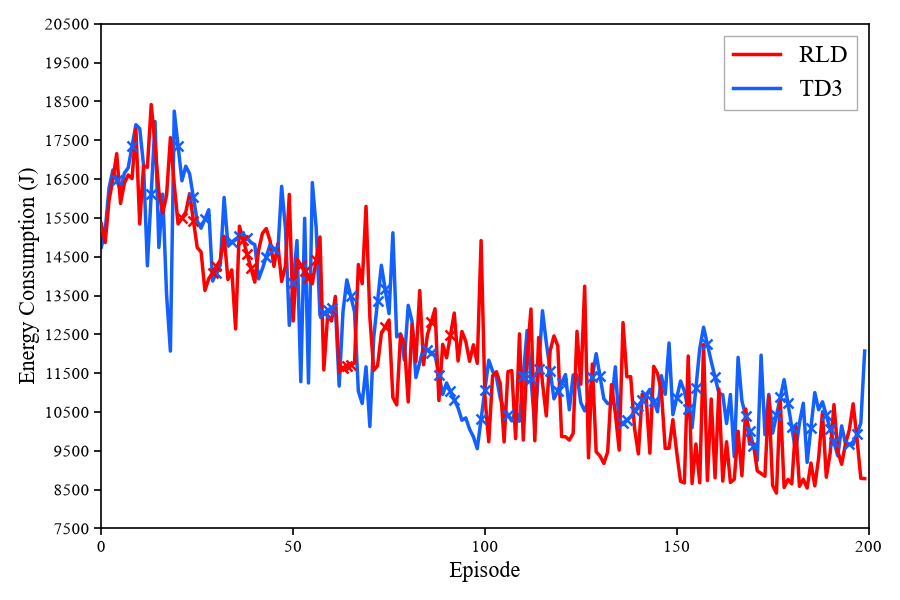}
    \caption{Energy consumption versus training episode for RLD and TD3.}
    \label{fig:convergence}
\end{figure}

Fig. \ref{fig:convergence} compares the energy consumption achieved by the proposed RLD scheme and the TD3 benchmark over 200 episodes in one training scenario. 
In the curves, the cross marks denote infeasible episodes, where the generated decisions fail to satisfy all constraints and therefore cannot complete the data processing tasks.
Both schemes exhibit a clear downward trend and gradually stabilize at lower levels as training proceeds.
The observed fluctuations mainly result from the use of different ID mobility scenarios during training, which leads to variations in channel conditions, task offloading feasibility, and energy consumption across episodes.
In the later episodes, RLD generally achieves lower energy consumption than TD3, with its values mostly distributed around $0.85\times10^4–1.05\times10^4$ J, whereas TD3 remains approximately within $0.95\times10^4–1.15\times10^4$ J, demonstrating the benefit of the proposed RLD algorithm in the corresponding optimization problem.

\begin{figure}[h]
    \centering
    \includegraphics[width=1\linewidth]{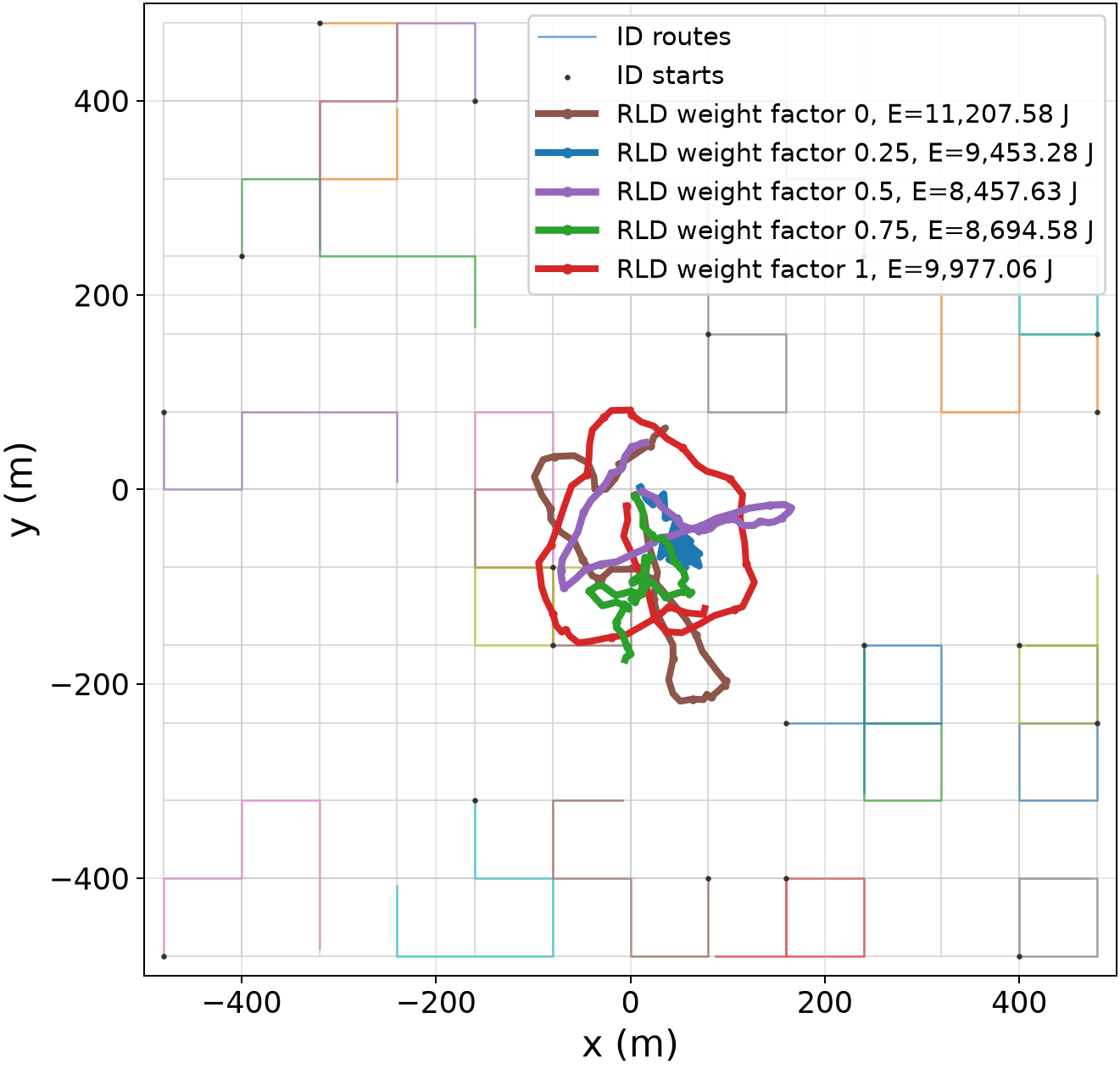}
    \caption{The UAV trajectories generated by different weight factor $\lambda_E$ in Scenario 1.}
    \label{weight_factor}
\end{figure}

We also evaluate the impact of the RAG weight factor $\lambda_E$ on the energy consumption of the proposed RLD algorithm. The results are shown in Fig. \ref{weight_factor}. It can be observed that RLD achieves the best performance when $\lambda_E$ is set to 0.5 or 0.75.
When $\lambda_E=0$, the RAG module retrieves historical cases only according to the similarity between the current ID mobility pattern and the historical ID paths. 
In this case, the historical energy performance is not considered during case retrieval. 
As a result, the retrieved UAV trajectories may only be similar to the current scenario but may not correspond to energy-efficient solutions, which weakens the guidance provided to the LLM and leads to higher energy consumption.
When $\lambda_E=1$, the RAG module overemphasizes historical cases with the best energy performance, while assigning insufficient importance to the similarity between the current ID mobility pattern and historical scenarios.
However, these cases may not match the current mobility pattern well, which reduces their usefulness as references for trajectory generation. 
Therefore, balancing scenario similarity and historical solution quality helps the RAG module provide more informative references for the LLM, thereby improving the performance of the RLD algorithm.

\section{Conclusions}
In this paper, we investigated energy-efficient task offloading and UAV trajectory planning in an MEC-empowered SAGIIN with mobile IDs and time-varying LEO satellite service windows. 
We formulated a joint optimization problem for UAV trajectory planning, task splitting, and computing resource allocation, and decomposed it into two subproblems. A RLD algorithm was proposed for UAV trajectory planning, while the task splitting and computing resource allocation subproblem was reformulated as a convex problem and solved using CVXPY. 
An energy-aware weighted retrieval mechanism was designed to balance mobility-pattern similarity and historical energy performance, providing informative guidance for DDPG exploration.
Simulation results demonstrated that RLD achieves lower energy consumption and maintains task feasibility across different mobility scenarios and ID densities. 
It also provides substantially lower online decision latency than conventional search-based algorithms.
The weight-factor analysis further validates the effectiveness of the proposed retrieval mechanism, highlighting the potential of RAG-assisted DRL for energy-efficient task offloading in dynamic SAGIINs.

\appendices
\renewcommand{\theequation}{\thesection--\arabic{equation}}

\ifCLASSOPTIONcaptionsoff
  \newpage
\fi

\bibliography{Reference}
\bibliographystyle{IEEEtran}  


\end{document}